\documentclass[aps,pra,reprint,groupedaddress,superscriptaddress]{revtex4-2}

\usepackage{amsmath}
\usepackage{amssymb}
\usepackage{graphicx}
\usepackage{braket}
\usepackage[colorlinks,linkcolor=blue, urlcolor=blue, anchorcolor=blue, citecolor=blue]{hyperref}

\begin{document}

\preprint{Research note}

\title{Optimal initial states for quantum Fisher information in linearized cavity optomechanics}

\author{Wangjun Lu}
\email{wjlu1227@zju.edu.cn}
\affiliation{School of Information Science and Engineering, Hunan Institute of Engineering, Xiangtan 411104, China}
\author{Qing Yu}
\affiliation{School of Information Science and Engineering, Hunan Institute of Engineering, Xiangtan 411104, China}
\author{Ying Li}
\affiliation{School of Information Science and Engineering, Hunan Institute of Engineering, Xiangtan 411104, China}
\author{Cuilu Zhai}
\affiliation{School of Information Science and Engineering, Hunan Institute of Engineering, Xiangtan 411104, China}
\author{Rui Zhang}
\affiliation{Key Laboratory of Low-Dimensional Quantum Structures and Quantum Control of Ministry of Education,
Synergetic Innovation Center for Quantum Effects and Applications,
XJ-Laboratory and Department of Physics, Hunan Normal University, Changsha 410081, China}
\author{Zhao-Hui Peng}
\affiliation{Hunan Provincial Key Laboratory of Intelligent Sensors and Advanced Sensor Materials, and Department of Physics, Hunan University of Science and Technology, Xiangtan 411201, China}
\author{Shiqing Tang}
\email{sqtang@hynu.edu.cn}
\affiliation{College of Physics and Electronic Engineering, Hengyang Normal University, Hengyang 421002, China}
\date{\today}

\begin{abstract}
We find the optimal initial states of a linearized cavity-optomechanical system for estimating the single-photon coupling by quantum Fisher information. In the red-detuned (beam-splitter) regime the interaction exchanges excitations between the two modes, and for pure inputs the Fisher information is four times the variance of the excitation-exchange operator, reducing the optimization to variance maximization with a closed-form solution. With the mechanical mode in a Fock state the probe is degenerate: any state saturating the photon-number constraint is optimal, and a Fock reference with many phonons enhances the per-photon sensitivity linearly. For an arbitrary fixed mechanical reference it becomes a phase-matched squeezed vacuum, from an exact bound on that coherence obtained by a Lagrange-dual argument. Jointly optimizing both modes at fixed total excitation number, the quantum Fisher information obeys a Heisenberg bound quadratic in the energy, saturated by equal superpositions of the two extremal eigenstates of the generator, equivalently a two-mode NOON state after a balanced beam splitter; balanced product Fock states reach only a linear enhancement. Exact diagonalization in truncated Fock spaces confirms all the results, the familiar single-photon input being the simplest saturating case. Three extensions are worked out: the blue sideband has an unbounded two-mode-squeezing spectrum, so the entangled optimum becomes impossible and the balanced product Fock state is exactly optimal at definite energy; under cavity loss the information accumulation turns linear on the cavity-lifetime timescale, at a renewal-optimal duration of about two and a half lifetimes; and for a state-of-the-art quantum-coherent coupling experiment the per-hour relative precision lies between one part in a million and ten parts in a million, number and thermal references being the most robust.
\end{abstract}

\maketitle

\section{\label{sec:intro}Introduction}

Cavity optomechanics couples the photon number of an optical cavity mode to the position of a mechanical oscillator. Since the first demonstrations of radiation-pressure cooling of micromechanical motion~\cite{Gigan2006,Arcizet2006}, anticipated theoretically in the quantum theory of cavity-assisted sideband cooling~\cite{WilsonRae2007,Marquardt2007} and extended to the resolved-sideband regime~\cite{Schliesser2008}, the field has grown into one of the most active platforms for quantum physics at the macroscale, as documented in the early mesoscale experiments~\cite{Kippenberg2008,Thompson2008} and in the standard reviews~\cite{Aspelmeyer2014,Poot2012}. Milestones along the way include quantum ground-state preparation of optical~\cite{Chan2011} and microwave~\cite{Teufel2011,Teufel2011b} resonators and, most recently, of Brillouin optomechanical cavities~\cite{Doeleman2023}, quantum-coherent beam-splitter-type exchange of excitations between light and mechanics~\cite{Verhagen2012} culminating in optomechanical quantum teleportation~\cite{Fiaschi2021}, coherent wavelength and microwave-to-optical conversion~\cite{Hill2012,Andrews2014,Bochmann2013,Bagci2014}, the direct observation of radiation-pressure quantum noise and of squeezed light generated by the mechanics itself~\cite{Purdy2013,SafaviNaeini2013,Wollman2015,Suh2014}, anticipated by proposals for parametric squeezing of the mirror motion~\cite{LiaoLaw2011squeezing}, real-time quantum feedback control of mechanical motion~\cite{Rossi2018,Magrini2018}, optomechanical entanglement in theory~\cite{Vitali2007,Genes2008} and experiment~\cite{Palomaki2013,Riedinger2018,OckeloenKorppi2018,Kotler2021,MercierdeLepinay2021}, mesoscopic superposition (cat) states of a massive mechanical oscillator~\cite{Bild2023}, and the single-photon strong-coupling regime~\cite{Rabl2011,Nunnenkamp2011}, alongside pulsed cooling~\cite{LiaoLaw2011chirped} and the spectroscopy of single-photon scattering~\cite{LiaoCheungLaw2012,LiaoLaw2013}.

A natural metrological question in this setting is how well the radiation-pressure coupling strength itself can be estimated from measurements on the outgoing light, and which initial quantum states of the two oscillators extract the most information. The ultimate precision of such an estimation is quantified by the quantum Fisher information (QFI) through the quantum Cram\'er--Rao bound~\cite{Helstrom1976,Braunstein1994}, an idea whose geometric roots lie in the statistical-distance interpretation of quantum states~\cite{Wootters1981} and whose modern role is codified in the general treatments of quantum metrology and quantum sensing~\cite{Giovannetti2004,Giovannetti2006,Boixo2007,Giovannetti2011,Paris2009,Degen2017}. It is by now well understood that nonclassical probe states beat classical ones. Squeezed states of light~\cite{Caves1981} reduce readout noise in precision interferometry~\cite{Caves1981} and have done so in real gravitational-wave detectors~\cite{Schnabel2010,Aasi2013,Tse2019,Acernese2019}, most recently with frequency-dependent squeezing in the Advanced LIGO detectors~\cite{Ganapathy2023}, and in quantum-enhanced searches for dark matter axions~\cite{Backes2021}. Correlated ions and spin-squeezed ensembles improve frequency standards~\cite{Wineland1994,Leibfried2005,Huelga1997} and enhance atom-interferometric sensors~\cite{Fuderer2023}; dual-Fock and large-NOON inputs reach the Heisenberg limit of interferometry~\cite{Holland1994,Boto2000,Nagata2007,Afek2010}; and two-mode squeezed states beat conventionally defined Heisenberg limits in specific protocols~\cite{Ou1992}. General bounds connect the QFI to collective spin observables and to the structure of multipartite entanglement~\cite{Pezze2009,Hyllus2012,Toth2012,Pezze2018}. For pure states undergoing unitary parameter encoding, the QFI reduces to four times the variance of the generator, which turns ``finding the best initial state'' into a variational problem with, in favorable cases, a closed-form answer; this strategy underlies the systematic optimization analyses of noisy optical interferometry~\cite{Dorner2009,Demkowicz2009}, of optimal Gaussian inputs for beam-splitter and active interferometers~\cite{Sparaciari2015}, and of Gaussian phase probes at fixed photon number~\cite{Monras2006}. In noisy environments the Heisenberg scaling is generally degraded~\cite{Escher2011,Demkowicz2012}, which motivates a careful treatment of the idealized (closed-system) optimum before adding loss.

For optomechanics specifically, the metrological spotlight has so far fallen mostly on optical phase, displacement, and force sensing, from back-action-evading and quantum-nondemolition strategies~\cite{Suh2014,MercierdeLepinay2021} to pulsed state-preparation schemes~\cite{Vanner2011}, spectrometric detection of weak forces in optomechanical cavities~\cite{ZhouLiao2020}, and feedback-based force tracking~\cite{Rossi2018,Magrini2018}, with parameter-estimation studies of related settings including gravimetry through nonlinear optomechanics~\cite{Qvarfort2018} and the estimation of quadratic position couplings~\cite{Wang2026}. The estimation of the single-photon coupling itself has received far less attention, even though it is the parameter that governs state exchange, transduction, and entanglement generation alike, and its precise value underlies the analysis of every experiment cited above. Closest to the present question, Bern\'ad, Sanavio, and Xuereb studied the Bayesian (minimum-mean-square-error) estimation of the nonlinear optomechanical coupling for a few-excitation optical input and a coherent mechanical state~\cite{Bernad2018}, with a dipole-coupling analog in Ref.~\cite{Bernad2019}, and Schneiter {et al.}\ computed the QFI for a time-modulated nonlinear coupling within the full nonlinear dynamics~\cite{Schneiter2020}. These works operate in the nonlinear regime, where no closed-form optimization over the initial states is available; in particular, none of them addresses the question we ask here: which initial states are optimal. This gap is worth closing: the single-photon coupling is simultaneously the figure of merit of the strong single-photon regime~\cite{Rabl2011,Nunnenkamp2011,LiaoLawKuangNori2015}, the calibration target of pulsed quantum optomechanics~\cite{Vanner2011}, and the rate-limiting quantity in quantum state exchange~\cite{Verhagen2012} and conversion~\cite{Hill2012,Andrews2014}. We close it here for the simplest and most experimentally relevant case: the linearized beam-splitter (red-sideband) interaction. Our central results are complete, closed-form solutions to two optimization problems that are natural inverses of each other: with the mechanical state fixed, find the optical state maximizing the quantum Fisher information for the coupling at a given energy, and with the optical state fixed, find the mechanical state maximizing the quantum Fisher information.
The answers exhibit a useful hierarchy. Number-state references make the quantum Fisher information a function of the probe energy only, with a degenerate optimum and a per-particle enhancement that grows linearly with the reference phonon number. Generic zero-mean references are handled by an exact bound on the two-photon coherence, which singles out phase-matched squeezed vacua as the unique optima. If both modes are jointly optimized at a fixed total energy, an entangled superposition of the extremal generator eigenstates saturates a Heisenberg bound that grows quadratically with the energy, and the single-photon input customarily used in optomechanical state-exchange schemes appears as its simplest instance. The framework is then extended in three directions, in Secs.~\ref{sec:blue}--\ref{sec:crb}: the counter-rotating (blue-sideband) interaction, whose generator has an unbounded spectrum, is solved in closed form; open-system dynamics under cavity loss and mechanical decoherence is analyzed, locating the crossover of the sensing precision from ideal quadratic-in-time growth to linear-in-time accumulation; and the whole hierarchy of strategies is benchmarked against the experimental parameters of the quantum-coherent coupling experiment of Ref.~\cite{Verhagen2012}.
The paper is organized as follows. Section~\ref{sec:model} introduces the model and the QFI framework and derives the master formula for the variance of the excitation-exchange operator. Sections~\ref{sec:optical} and~\ref{sec:mechanical} solve these two problems in turn. Section~\ref{sec:joint} treats jointly optimized inputs at definite energy on the red sideband, and Sec.~\ref{sec:blue} solves the blue-sideband (two-mode-squeezing) variant. Section~\ref{sec:open} extends the analysis to open-system dynamics and locates the transition from quadratic to linear information accumulation. Section~\ref{sec:crb} collects the consequences for estimating $g_0$ and benchmarks the strategies against experimental parameters. Section~\ref{sec:conclusion} concludes. All derivations are given in the main text, immediately following the results they establish.

\section{\label{sec:model}Model and quantum Fisher information}

\subsection{\label{sec:ham}Linearized optomechanical interaction}

\begin{figure}[t]
\includegraphics[width=\columnwidth]{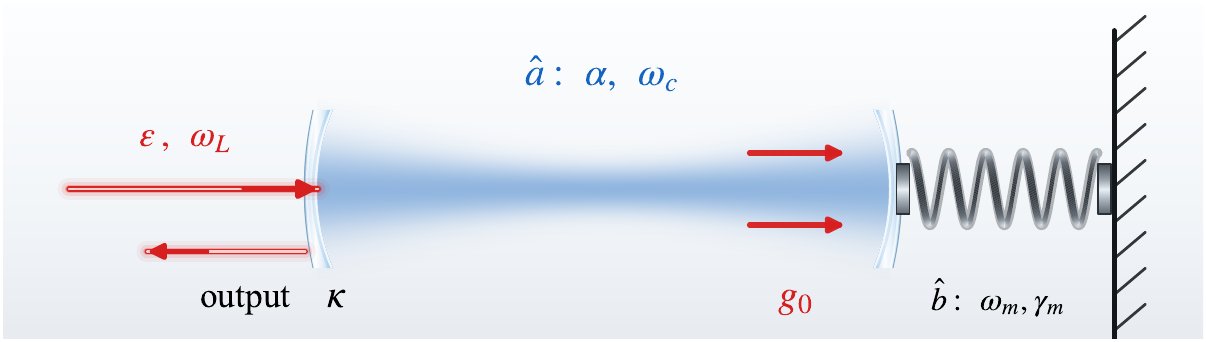}
\caption{\label{fig:setup}Schematic of the model. A driven Fabry--P\'erot
cavity supports a single optical mode of resonance frequency $\omega_c$ and
energy-decay rate $\kappa$; its right mirror is a mechanical oscillator of
frequency $\omega_m$ and damping rate $\gamma_m$, whose displacement $x$ is
quantized through the phonon annihilation operator $\hat b$. Radiation
pressure couples the intracavity photon number to this displacement with
single-photon coupling $g_0$ (red arrows). A coherent drive of frequency
$\omega_L$ and amplitude $\varepsilon$ populates the cavity with a large
classical amplitude $\alpha$, and the output field leaking through the input
mirror carries the information on $g_0$. In the rotating frame of the drive
the laser--cavity detuning is $\Delta=\omega_L-\omega_c$; on the red
mechanical sideband, $\Delta=-\omega_m$, the linearized coupling reduces to
the beam-splitter Hamiltonian~(\ref{eq:hint}).}
\end{figure}

The single-mode optomechanical Hamiltonian in a rotating frame at the laser frequency $\omega_L$ reads~\cite{Aspelmeyer2014}
\begin{equation}
\label{eq:hamiltonian}
\hat H = -\hbar\Delta\, \hat a^\dagger \hat a + \hbar\omega_m \hat b^\dagger \hat b
-\hbar g_0\, \hat a^\dagger \hat a\,(\hat b+\hat b^\dagger)
+ i\hbar\varepsilon\,(\hat a^\dagger-\hat a),
\end{equation}
where $\hat a$ ($\hat b$) is the annihilation operator of the optical (mechanical) mode, of resonance frequency $\omega_c$ ($\omega_m$); $\Delta=\omega_L-\omega_c$ is the laser--cavity detuning; $g_0$ is the single-photon radiation-pressure coupling; $\varepsilon$ is the coherent drive amplitude; and $\kappa$ denotes the cavity energy-decay rate, which enters the steady-state intracavity amplitude below. The passage from Eq.~(\ref{eq:hamiltonian}) to the effective interaction used throughout this paper proceeds in three steps, which we now describe in turn. The physical setting is sketched in Fig.~\ref{fig:setup}.

The first step is the displacement of the optical mode. For a strong drive the field operator is written as the sum of a classical coherent amplitude and a quantum fluctuation, $\hat a=\alpha+\hat d$, where $\alpha=\varepsilon/(\kappa/2-i\Delta)$ is the steady-state intracavity amplitude, independent of $g_0$ to leading order, and $\hat d$ is the annihilation operator of the optical fluctuation. Substituting into Eq.~(\ref{eq:hamiltonian}) and using $\hat a^\dagger\hat a=|\alpha|^2+\alpha^*\hat d+\alpha \hat d^\dagger+\hat d^\dagger\hat d$, the radiation-pressure term separates into three pieces,
\begin{align}
-\hbar g_0 \hat a^\dagger \hat a(\hat b+\hat b^\dagger)
&=-\hbar g_0|\alpha|^2(\hat b+\hat b^\dagger)\nonumber\\
&\quad-\hbar G\,(\hat d+\hat d^\dagger)(\hat b+\hat b^\dagger)
-\hbar g_0\, \hat d^\dagger \hat d\,(\hat b+\hat b^\dagger),
\label{eq:expand}
\end{align}
with $G\equiv g_0\alpha$. The three terms on the right-hand side play distinct physical roles. The first term is a $c$-number force, proportional to the coherent photon number $|\alpha|^2$, which merely shifts the mechanical equilibrium position and can be absorbed into a redefinition of $\hat b$. The second term is linear in the fluctuation amplitudes of both modes and is enhanced by the coherent amplitude, $|\alpha|\gg1$; this coherent amplification is what makes the linearized theory accurate even though $g_0$ itself is tiny. The third term, quadratic in the optical fluctuation, carries only a bare factor $g_0$ without enhancement and is negligible against the linear term. Keeping the linear term alone leaves the quadratic Hamiltonian
\begin{equation}
\label{eq:hlin}
\hat H_{\rm lin}=-\hbar\Delta\, \hat d^\dagger \hat d+\hbar\omega_m \hat b^\dagger \hat b
-\hbar G\,(\hat d+\hat d^\dagger)(\hat b+\hat b^\dagger).
\end{equation}

The second step is the passage to the interaction picture with respect to the free Hamiltonian $\hat H_0=-\hbar\Delta\, \hat d^\dagger \hat d+\hbar\omega_m \hat b^\dagger \hat b$. Under this free evolution each product operator acquires a phase factor at the sum or the difference of the detuning and the mechanical frequency, and the coupling operator in Eq.~(\ref{eq:hlin}) expands into four terms,
\begin{equation}
\label{eq:fourterms}
(\hat d+\hat d^\dagger)(\hat b+\hat b^\dagger)\;\longrightarrow\;
\hat d\,\hat b^\dagger+\hat d^\dagger \hat b
+\hat d\,\hat b+\hat d^\dagger \hat b^\dagger ,
\end{equation}
which rotate as $e^{i(\Delta+\omega_m)t}$, $e^{-i(\Delta+\omega_m)t}$, $e^{i(\Delta-\omega_m)t}$, and $e^{-i(\Delta-\omega_m)t}$, respectively: the first two terms (one phonon created, one fluctuation photon annihilated, and its time reverse) oscillate at the sum frequency $\Delta+\omega_m$, while the last two (simultaneous creation or annihilation of one quantum in each mode) oscillate at the difference frequency $\Delta-\omega_m$.

The third step is the rotating-wave approximation. In the resolved-sideband regime and on the red mechanical sideband, $\Delta=-\omega_m$, the first two terms in Eq.~(\ref{eq:fourterms}) are resonant while the last two rotate at twice the mechanical frequency, $e^{\mp 2i\omega_m t}$; for a weak coupling $G\ll\omega_m$ their effect averages out over the interaction time and they are discarded. This leaves the beam-splitter interaction
\begin{equation}
\label{eq:hint}
\hat H_{\rm int}=\hbar G\,\hat n,\qquad
\hat n\equiv \hat d\,\hat b^\dagger+\hat d^\dagger \hat b,\qquad
G\equiv g_0\alpha ,
\end{equation}
i.e.\ on the red sideband each term of the drive-induced linear coupling describes the annihilation of one fluctuation photon and the creation of one phonon, or the time reverse, an ideal excitation-exchange channel. (On the blue sideband, $\Delta=+\omega_m$, the RWA instead retains the two-mode-squeezing terms $\hat d\,\hat b+\hat d^\dagger \hat b^\dagger$; that case is analyzed in Sec.~\ref{sec:blue}.)
This is the interaction that implements photon--phonon state exchange~\cite{Verhagen2012} and wavelength conversion~\cite{Hill2012,Andrews2014}. We emphasize the division of labor: the drive amplitude $\alpha$ is a controllable classical quantity, while $g_0$ is the parameter to be estimated; the information about $g_0$ is carried exclusively through the dimensionless angle $\theta\equiv Gt=g_0\alpha t$ in the unitary
\begin{equation}
\label{eq:unitary}
\hat U(\theta)=\exp(-i\theta\,\hat n).
\end{equation}

\subsection{\label{sec:qfi}QFI for fixed-generator unitary encoding}

For a pure input state $\ket{\psi_0}$ and the encoding~(\ref{eq:unitary}), the QFI with respect to $\theta$ is~\cite{Braunstein1994}
\begin{equation}
\label{eq:qfivar}
F_Q^{(\theta)} = 4\left(\braket{\partial_\theta\psi|\partial_\theta\psi}
-\big|\braket{\psi|\partial_\theta\psi}\big|^2\right)
= 4\,\mathrm{Var}_{\psi_0}(\hat n),
\end{equation}
where $\mathrm{Var}_{\psi_0}(\hat n)=\braket{\hat n^2}-\braket{\hat n}^2$ is the variance of $\hat n$ evaluated on the initial state. This reduction deserves a word of explanation, because it is what makes the entire optimization tractable. Differentiating the encoded state gives $\ket{\partial_\theta\psi}=-i\hat n\,\hat U(\theta)\ket{\psi_0}$, and since the generator $\hat n$ itself is parameter-independent, it commutes with the encoding unitary; the two terms in the QFI therefore involve only matrix elements of $\hat n$ between the input state and its image, which combine into the variance of $\hat n$ evaluated on $\ket{\psi_0}$. Physically, the parameter is imprinted on the state as a rotation on the unit sphere of the operator $\hat n$, and the distinguishability of infinitesimally close rotations is governed entirely by how spread out the input state is over the eigenstates of $\hat n$: an eigenstate of $\hat n$ acquires only a global phase and carries no information, while a superposition of two eigenstates with maximal eigenvalue separation is maximally sensitive. Equivalently, $\ket{\psi_0}$ and $\hat U(\theta)\ket{\psi_0}$ have the same spectral decomposition of $\hat n$ probabilities, so no property of the output state beyond the input-state variance matters.

The QFI for $\theta$ translates into the QFI for the physical coupling $g_0$ by the chain rule of classical Fisher information theory, which applies unchanged to the quantum quantity. Since $\partial\theta/\partial g_0=\alpha t$ at fixed $\alpha$ and $t$, one finds
\begin{equation}
\label{eq:chain}
F_Q^{(g_0)}=(\alpha t)^2\,F_Q^{(\theta)}.
\end{equation}
the superscript labelling the parameter being estimated, the rotation angle $\theta$ or the physical coupling $g_0$. Estimating $g_0$ is thus equivalent to estimating the rotation angle $\theta$, with a trivial prefactor: whatever state optimizes one optimizes the other. Equation~(\ref{eq:chain}) calls for two comments. The prefactor is quadratic in the interaction time $t$, reflecting the coherent accumulation of the rotation angle; this quadratic-in-time scaling is the signature of an ideal, loss-free channel and will be revisited when open-system effects are discussed in Sec.~\ref{sec:open}. The prefactor is also proportional to the coherent enhancement $|\alpha|$: although the parameter $g_0$ multiplies the operator $\hat a^\dagger\hat a(\hat b+\hat b^\dagger)$, which does not conserve anything useful, the strong drive converts it into a rotation generated by the conserved operator $\hat n$ at an amplified rate $G=g_0\alpha$, which is why the linearized problem has clean symmetries that the bare radiation-pressure problem lacks. Throughout, $N\equiv\braket{\hat n_d}$ and $M\equiv\braket{\hat n_b}$ denote the mean excitation numbers of the input state, with $\hat n_d\equiv \hat d^\dagger\hat d$ and $\hat n_b\equiv \hat b^\dagger \hat b$ the number operators of the optical and the mechanical fluctuation mode.

\subsection{\label{sec:schwinger}Schwinger representation and master formula}

The operator $\hat n$ has two structural properties that organize everything that follows. The first is the conservation law: since $\hat n$ exchanges one excitation between the modes, it commutes with the total excitation number operator $\hat n_d+\hat n_b$ and therefore conserves it. The second is the algebraic identity with the Schwinger representation of angular momentum: defining
\begin{equation}
\hat J_z=\tfrac{1}{2}(\hat d^\dagger \hat d-\hat b^\dagger \hat b),\qquad
\hat J_x=\tfrac{1}{2}(\hat d\,\hat b^\dagger+\hat d^\dagger \hat b)=\tfrac12\hat n,
\end{equation}
one checks that $\hat J_x$, $\hat J_y=\tfrac{1}{2i}(\hat d\,\hat b^\dagger-\hat d^\dagger \hat b)$, and $\hat J_z$ obey the $su(2)$ algebra with total angular momentum $\hat J^2$ proportional to the conserved quantity. In other words, $\hat n=2\hat J_x$ is twice the $x$ component of a Schwinger pseudospin, and the parameter $\theta$ generates a rotation of this pseudospin about the $x$ axis.

The conservation of the total excitation number, $\braket{\hat n_d+\hat n_b}=E$, implies that the Hilbert space decomposes into a direct sum of $SU(2)$ manifolds of dimension $E+1$, spanned by $\{\ket{k,E-k}\}_{k=0}^{E}$, with each manifold carrying one irreducible spin-$E/2$ representation. On each manifold the generator $\hat n=2\hat J_x$ has the spectrum $\{-E,-E+2,\dots,E\}$: the eigenvalues are equally spaced by two, the extremal eigenvalues are attained on unique eigenstates $\ket{\pm}$, and the manifold label $E$ simultaneously sets the maximal variance that any state on it can reach. This geometric structure underlies both the moment bound of Sec.~\ref{sec:generalref} and the Heisenberg bound of Sec.~\ref{sec:joint}.

For a product input $\ket{\psi}_d\otimes\ket{\phi}_b$ a direct expansion gives the master formula
\begin{align}
\label{eq:master}
\mathrm{Var}(\hat n) &= 2\,\mathrm{Re}\!\left[\braket{\hat d^2}\braket{\hat b^{\dagger 2}}\right]
+2NM+N+M \nonumber\\[2pt]
&\quad -4\left(\mathrm{Re}\!\left[\braket{\hat d}\braket{\hat b}^{*}\right]\right)^2 ,
\end{align}
where every expectation value is taken in the initial state, factorized over the two modes: $\braket{\hat d}$, $\braket{\hat d^2}$, and $N=\braket{\hat n_d}$ are evaluated in the optical factor $\ket{\psi}_d$, while $\braket{\hat b}$, $\braket{\hat b^{\dagger2}}$, and $M=\braket{\hat n_b}$ are evaluated in the mechanical factor $\ket{\phi}_b$. No expectation value involves the joint state or the evolution: the formula relates the four initial-state moments of the two modes, and this is the sense in which the optimization of the QFI is a problem about initial states only. Every optimization below is a consequence of this single equation. Three structural facts are at work. The term $2NM+N+M$ depends only on the mean occupations and forms a state-independent baseline once the energies are fixed; pair coherences couple through $\braket{\hat d^2}\braket{\hat b^{\dagger2}}$, rewarding phase-matched squeezing on both sides, and this is the only term that can depend on the squeezing phases, with its sign controlled by their relative phase; and coherent amplitudes enter only through the nonpositive term $-4\,\mathrm{Re}^2[\braket{\hat d}\braket{\hat b}^*]$, so that phase-aligning the two coherent amplitudes can only reduce the QFI, consistent with the intuition that a shared phase-space displacement carries no information about $\theta$; a common displacement of both modes is a symmetry of the interaction, and the state can be made to ``ride along'' with the rotation.

\subsection{\label{app:master}Derivation of the master formula}

We record the derivation of Eq.~(\ref{eq:master}) for completeness. The variance is $\mathrm{Var}(\hat n)=\braket{\hat n^2}-\braket{\hat n}^2$, so we need $\braket{\hat n^2}$ and $\braket{\hat n}$ on a general product input. Squaring the generator and using the bosonic commutation relations with care for operator ordering gives
\begin{equation}
\hat n^2=\hat d^2\hat b^{\dagger2}+\hat d^{\dagger2}\hat b^2+(\hat n_d+1)\hat n_b+\hat n_d(\hat n_b+1),
\end{equation}
where we have used $\hat d\,\hat b^\dagger \hat d\,\hat b^\dagger=\hat d^2\hat b^{\dagger2}$ (the operators acting on different modes commute, so the two creation operators of $\hat b$ collected from the two factors of $\hat n$ combine into $\hat b^{\dagger2}$), and $\hat d\,\hat b^\dagger \hat d^\dagger \hat b=(\hat n_d+1)\hat n_b$ as well as $\hat d^\dagger \hat b\,\hat d\,\hat b^\dagger=\hat n_d(\hat n_b+1)$, both of which follow from $\hat d\,\hat d^\dagger=\hat n_d+1$ and the cross-mode commutativity. Taking expectation values, all cross-mode operator products factorize across the product state into a product of single-mode moments,
\begin{equation}
\braket{\hat n^2}
=2\,\mathrm{Re}\!\left[\braket{\hat d^2}\braket{\hat b^{\dagger2}}\right]
+(N+1)M+N(M+1),
\end{equation}
where the reality of the two number terms has been used to combine them, while the mean value of the generator is simply $\braket{\hat n}=2\,\mathrm{Re}[\braket{\hat d}\braket{\hat b}^*]$. The two modes enter the mean only through their first moments, and through the product $\braket{\hat d}\braket{\hat b}^*$; squaring this expression and subtracting from $\braket{\hat n^2}$ gives precisely Eq.~(\ref{eq:master}). We emphasize that nowhere in the derivation did we use any property of the two states beyond their being product states: the master formula holds for arbitrary, including non-Gaussian and highly entangled-in-single-mode, inputs of the two factors.

\section{\label{sec:optical}Fixed mechanical state: optimal optical probe}

\subsection{\label{sec:numberref}Number-state reference: degenerate optimum and $(2m+1)$ enhancement}

\begin{table*}[t]
\caption{\label{tab:table1}QFI density $F_Q^{(\theta)}/4=\mathrm{Var}(\hat n)$ for product inputs at mean occupations $(N,M)$, Eq.~(\ref{eq:master}). $\gamma$ ($\beta$) denotes the coherent amplitude of the optical (mechanical) mode. ``Fock ref.''\,=\,mechanical Fock state $\ket{m}$ with $M=m$; ``coherent ref.''\,=\,mechanical coherent state $\ket{\beta}$ with $M=|\beta|^2$; ``squeezed ref.''\,=\,mechanical squeezed vacuum with $M=\sinh^2 r_b$ and $\braket{\hat b^2}=\sqrt{M(M+1)}\,e^{i\varphi_b}$. The probe squeezed states carry $\braket{\hat d^2}=\sqrt{N(N+1)}\,e^{i\varphi_a}$; $\varphi\equiv\varphi_a+\varphi_b$ is the squeezing-phase mismatch, and the maximal values assume optimal phase matching, Eq.~(\ref{eq:phasematch}).}
\begin{ruledtabular}
\scriptsize
\begin{tabular}{lccc}
 & Fock ref. $\ket{m}$ & coherent ref. $\ket{\beta}$ & squeezed ref. \\
\hline
Fock probe $\ket{N}$ & $(2m+1)N+m$ & $(2M+1)N+M$ & $(2M+1)N+M$ \\
coherent probe $\ket{\gamma}$, $N=|\gamma|^2$ & $(2m+1)N+m$ & \begin{tabular}{l}$(2M+1)N+M$\\[-1pt]${+2\,\mathrm{Re}[\gamma^2\beta^{*2}]-4\,\mathrm{Re}^2[\gamma\beta^*]}$\end{tabular} & $(2M+1)N+M+2N\sqrt{M(M+1)}\cos\varphi$ \\
squeezed probe, $N=\sinh^2 r_a$ & $(2m+1)N+m$ & $(2M+1)N+M+2M\sqrt{N(N+1)}\cos\varphi$ & \begin{tabular}{l}$(2M+1)N+M$\\[-1pt]${+2\sqrt{N(N+1)M(M+1)}\cos\varphi}$\end{tabular} \\
\end{tabular}
\end{ruledtabular}
\end{table*}

\begin{figure}[t]
\includegraphics[width=\columnwidth]{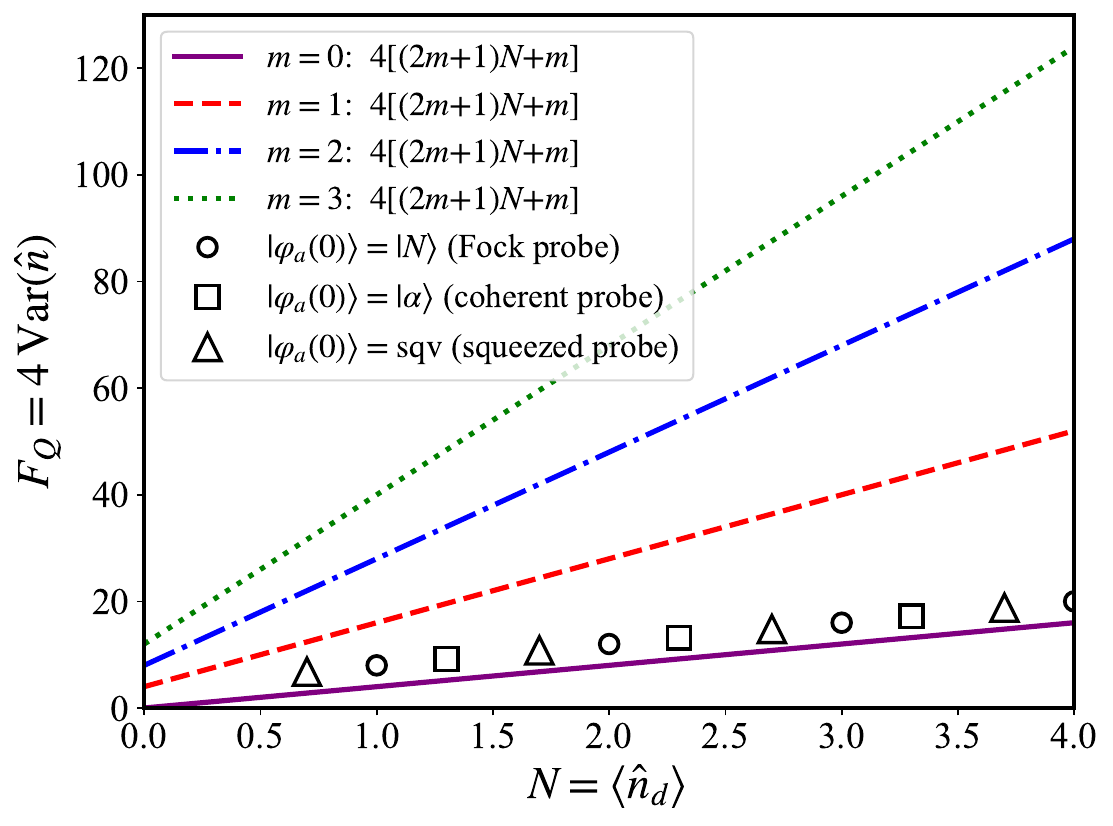}
\caption{\label{fig:numref}Number-state reference (Theorem~1). $F_Q^{(\theta)}/4=\mathrm{Var}(\hat n)$ as a function of the probe mean photon number $N$ for Fock references $m=0,1,2,3$: straight lines of slope $2m+1$. Markers show Fock, coherent, and squeezed-vacuum probes of equal mean photon number on the $m=1$ line; all probe families of equal energy perform identically, so the optimum is degenerate.}
\end{figure}

Suppose the mechanical mode is prepared in the Fock state $\ket{m}$. Then $\braket{\hat b}=\braket{\hat b^2}=0$, and the master formula~(\ref{eq:master}) collapses to
\begin{equation}
\label{eq:theorem1}
F_Q^{(\theta)}=4\big[(2m+1)\braket{\hat n_d}+m\big]
\qquad\text{(any optical state)}.
\end{equation}
The QFI depends on the optical state only through its mean photon number: all optical states with the same energy are exactly equivalent. The statement is stronger than it looks. It holds for arbitrary, non-Gaussian, arbitrarily structured optical states, and it is a direct consequence of the structure of the master formula: the Fock reference has neither a coherent amplitude nor pair coherence, so the only surviving link between the optical state and the master formula is the single moment $\braket{\hat n_d}$. Under the energy constraint $\braket{\hat n_d}\le N$ the optimum is therefore degenerate: any state saturating the constraint is optimal, and
\begin{equation}
\label{eq:FQg0}
F_Q^{(g_0)}=4(\alpha t)^2\big[(2m+1)N+m\big].
\end{equation}
The linear scaling is illustrated in Fig.~\ref{fig:numref}. Two remarks follow. For $m=0$ this reproduces the familiar single-photon result: the input $\ket{1}_d\ket{0}_b$, one fluctuation photon and no phonon, gives $F_Q^{(\theta)}=4$ independent of the interaction time, with Cram\'er--Rao bound $\Delta g_0\ge 1/(2\alpha t\sqrt{\nu})$ for $\nu$ repetitions. And each additional phonon in the reference enhances the per-photon sensitivity by the factor $2m+1$: the number state acts as an ideal ``force reference'' whose displacement by a single fluctuation quantum is maximally distinguishable. The enhancement is easy to read off from the Schwinger picture: a Fock reference with $m$ phonons places the joint state on the $SU(2)$ manifold with total excitation $E=n+m$, where the generator spectrum has spacing two and is centered $m$ quanta away from vacuum, so every optical quantum rotates against a stiffer pseudospin. The degeneracy of the optimum is a peculiarity of the number reference (and of the beam-splitter interaction): since $\ket{m}$ has no coherent or pair coherence, the probe's $\braket{\hat d^2}$ and $\braket{\hat d}$ enter the master formula only through terms that vanish.

By the interchange symmetry $\hat d\leftrightarrow \hat b$ of $\hat n$, the identical statement holds with the roles exchanged: an optical Fock state $\ket{n}$ as reference gives $F_Q^{(\theta)}=4[(2n+1)\braket{\hat n_b}+n]$ for arbitrary mechanical states; see Sec.~\ref{sec:mechanical}.

\subsection{\label{sec:generalref}General fixed reference: phase-matched squeezed probe}

\begin{figure}[t]
\includegraphics[width=\columnwidth]{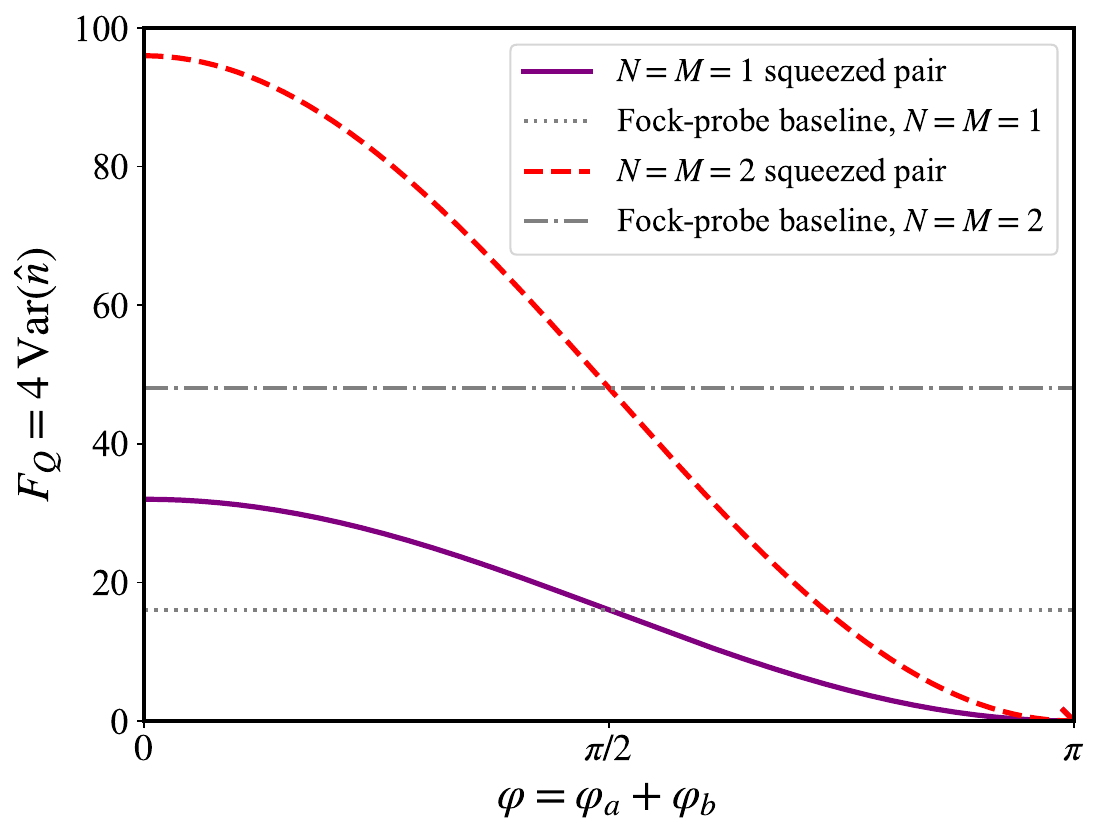}
\caption{\label{fig:phase}Cost of squeezing-phase mismatch: $F_Q^{(\theta)}/4$ versus the mismatch angle $\varphi=\varphi_a+\varphi_b$ for phase-matched squeezed probe--reference pairs at $N=M=1$ (blue) and $N=M=2$ (red). $\varphi=0$: optimum; $\varphi=\pi/2$: the pair-coherence term vanishes; $\varphi=\pi$: it flips sign and the QFI drops below the phase-insensitive Fock-probe baseline (dotted horizontal lines, $\mathrm{Var}= (2M{+}1)N{+}M$). Crosses mark the $\varphi=\pi$ values.}
\end{figure}

\begin{figure}[t]
\includegraphics[width=\columnwidth]{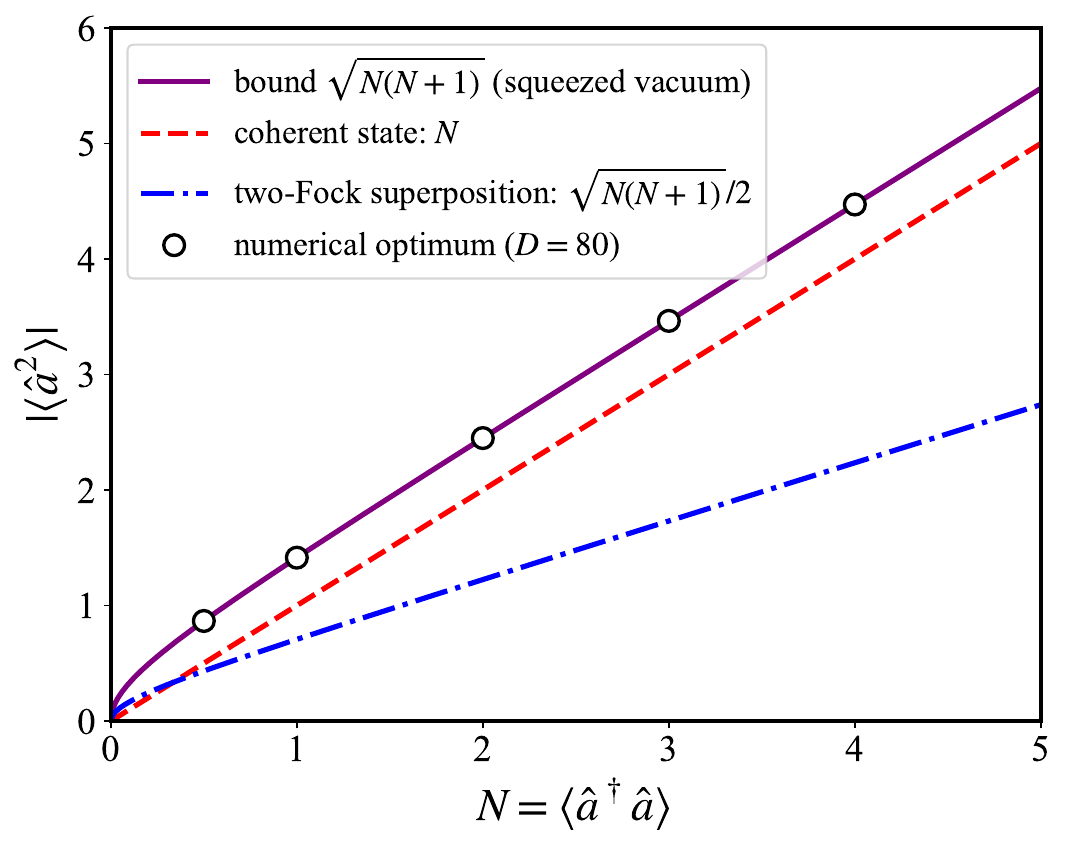}
\caption{\label{fig:moment}The exact moment bound~(\ref{eq:momentbound}). Solid green: $|\braket{\hat a^2}|=\sqrt{N(N+1)}$, saturated uniquely by the squeezed vacuum; dashed grey: coherent states ($|\braket{\hat a^2}|=N$); dotted blue: two-Fock superpositions ($\sqrt{N(N+1)}/2$). Circles: numerical optima obtained by constrained maximization over truncated Fock space ($D=80$, Lagrange-dual bisection), lying on the bound.}
\end{figure}

We now let the mechanical mode be fixed in an arbitrary pure state $\ket{\phi}$ and optimize over the optical probe at fixed mean photon number $\braket{\hat n_d}=N$. Writing $\beta=\braket{\hat b}$ for the coherent amplitude, $B=\braket{\hat b^{\dagger2}}$ for the pair coherence of the reference, and $M=\braket{\hat n_b}$ for its mean occupation, the master formula~(\ref{eq:master}) takes the form
\begin{equation}
\label{eq:varfixedref}
\mathrm{Var}(\hat n)=2NM+N+M
+2\,\mathrm{Re}\!\left[\braket{\hat d^2}B\right]
-4\,\mathrm{Re}^2\!\left[\braket{\hat d}\beta^*\right].
\end{equation}
The first term $2NM+N+M$ is a fixed baseline: it depends only on the mean occupations of the two modes and cannot be tuned by the choice of probe. The last term is nonpositive and is minimized to zero by any probe with $\braket{\hat d}=0$, or more generally by any probe whose coherent amplitude is in phase quadrature with that of the reference, $\mathrm{Re}[\braket{\hat d}\beta^*]=0$; in either case this term can be eliminated entirely, so the only quantity left to optimize is the pair-coherence term $2\,\mathrm{Re}[\braket{\hat d^2}B]$. That term rewards an optical probe with a large two-photon coherence $\braket{\hat d^2}$ whose phase is aligned with the complex conjugate of the reference's pair coherence, and its maximization is limited by an exact moment bound proved below (see Fig.~\ref{fig:moment} for a comparison of the bound with competing state families and with numerical optima):
\begin{equation}
\label{eq:momentbound}
\big|\braket{\hat a^2}\big|\;\le\;\sqrt{N(N+1)}
\qquad\text{at fixed }\braket{\hat a^\dagger \hat a}=N,
\end{equation}
with equality attained uniquely (up to a phase) by the squeezed vacuum. [For comparison, coherent states reach only $|\braket{\hat a^2}|=N$, and two-component Fock superpositions $(N-1)/2$.] Consequently,
\begin{equation}
\label{eq:theorem3}
F_Q^{(\theta)}\le 4\big[(2M+1)N+M+2\sqrt{N(N+1)}\,|B|\big],
\end{equation}
and equality holds iff the optical probe is a squeezed vacuum whose squeezing axis is matched to the reference pair coherence,
\begin{equation}
\label{eq:phasematch}
\arg\braket{\hat d^2}=-\arg\braket{\hat b^{\dagger2}}
\qquad\text{(phase matching)},
\end{equation}
with $\braket{\hat d}=0$ (or, more generally, $\mathrm{Re}[\braket{\hat d}\beta^*]=0$). Denoting the probe and the reference squeezing phases by $\varphi_a=\arg\braket{\hat d^2}$ and $\varphi_b=-\arg\braket{\hat b^{\dagger2}}$, only their sum $\varphi\equiv\varphi_a+\varphi_b$ enters the QFI, so that phase matching means $\varphi=0$. The phase-matching condition has a direct physical meaning. A squeezed vacuum probes the generator with a two-photon coherence whose phase is free; the reference contributes its own pair coherence with a fixed phase; the QFI depends on the cosine of the relative phase. Aligning the two axes makes the two-photon coherences add constructively, misalignment reduces the gain continuously, and complete misalignment ($\varphi=\pi$) is worse than using no squeezing at all, because the pair-coherence term then actively subtracts from the baseline. Table~\ref{tab:table1} collects closed-form results for the standard Gaussian and number-state combinations, including the phase dependence $\cos\varphi$ of misaligned squeezing; Fig.~\ref{fig:phase} shows how much the phase matters: for $N=M=1$ and the reference $(\ket{0}+\ket{2})/\sqrt{2}$ the optimal probe yields $F_Q^{(\theta)}=24$ while the $\pi$-misaligned probe yields only $8$, i.e., $2\,\mathrm{Re}[\braket{\hat d^2}\braket{\hat b^{\dagger2}}]=\pm2$ instead of $+2$ on top of the baseline $2NM+N+M=4$.

We note that the bound~(\ref{eq:momentbound}) is the optomechanical analog of the statement that squeezed vacuum is the optimal pure Gaussian probe for optical phase estimation at fixed photon number~\cite{Monras2006}; here it holds globally over all states, Gaussian or not, and it is the ingredient that makes Theorem~3 exact rather than merely Gaussian.

\subsection{\label{app:moment}Proof of the moment bound $\max|\braket{\hat a^2}|=\sqrt{N(N+1)}$}

We prove Eq.~(\ref{eq:momentbound}). The proof strategy is a Lagrange-dual argument: instead of maximizing $|\braket{\hat a^2}|$ directly under the energy constraint (a nonconvex problem over states, on which numerical optimizers routinely stall), we construct a family of operator upper bounds on $\mathrm{Re}\braket{\hat a^2}$ parametrized by a free multiplier $\lambda$, and then minimize the bound over $\lambda$. The best bound in the family will turn out to be exactly $\sqrt{N(N+1)}$, and the state saturating it will be identified explicitly, which upgrades the dual statement to an exact global optimum.

By a phase rotation of $\hat a$ it suffices to bound $\mathrm{Re}\braket{\hat a^2}$, since $|\braket{\hat a^2}|$ equals its real part in a suitably rotated frame and both the constraint $\braket{\hat a^\dagger \hat a}=N$ and the set of attainable states are invariant under such rotations. For any unit vector $\ket{\psi}$ and any $\lambda\ge2$,
\begin{equation}
\mathrm{Re}\braket{\hat a^2}\le\frac12\braket{\hat a^2+\hat a^{\dagger2}}
=\frac12\Big(\braket{\hat H_\lambda}+\lambda\braket{\hat n}\Big),
\end{equation}
with $\hat H_\lambda\equiv \hat a^2+\hat a^{\dagger2}-\lambda\,\hat a^\dagger \hat a$: this is simply the statement that the expectation of $\hat a^2+\hat a^{\dagger2}$ in any state is a real number bounded by its absolute value, rewritten in terms of $\hat H_\lambda$. The next step is to evaluate $\sup_\psi\braket{\hat H_\lambda}$ exactly. The operator $\hat H_\lambda$ is quadratic, so a Bogoliubov rotation diagonalizes it: write
\begin{equation}
\hat a=\cosh r\;\hat c-\sinh r\;\hat c^\dagger
\end{equation}
(a canonical transformation, $[\hat c,\hat c^\dagger]=1$, for every $r$). Expanding in terms of $\hat c$ and $\hat c^\dagger$,
\begin{align}
\hat H_\lambda&=\Big(\cosh2r+\tfrac{\lambda}{2}\sinh2r\Big)
\big(\hat c^2+\hat c^{\dagger2}\big)\nonumber\\
&\quad+\big({-2\sinh2r-\lambda\cosh2r}\big)\,\hat c^\dagger \hat c
\;-\sinh2r-\lambda\sinh^2r .
\label{eq:bogoliubov}
\end{align}
The first term is the unwanted non-number-conserving piece. Choosing $\tanh2r=-2/\lambda$ (possible for $\lambda\ge2$) eliminates it; then
$\sinh2r=-2/\sqrt{\lambda^2-4}$, $\cosh2r=\lambda/\sqrt{\lambda^2-4}$, and
\begin{equation}
\hat H_\lambda=-\sqrt{\lambda^2-4}\;\hat c^\dagger \hat c
+\frac{\lambda-\sqrt{\lambda^2-4}}{2},
\end{equation}
whose spectrum is bounded above by $\big(\lambda-\sqrt{\lambda^2-4}\big)/2$, attained on the $\hat c$-vacuum. The diagonalization has converted the constrained operator bound into an explicit constant. Hence, for every $\lambda\ge2$ and every state with $\braket{\hat n}=N$,
\begin{equation}
\mathrm{Re}\braket{\hat a^2}\le\frac12\left[\lambda N
+\frac{\lambda-\sqrt{\lambda^2-4}}{2}\right].
\end{equation}
Minimizing the right-hand side over $\lambda$ gives
$\lambda_\ast=(2N+1)/\sqrt{N(N+1)}\ \ge2$, and
$\sqrt{\lambda_\ast^2-4}=1/\sqrt{N(N+1)}$, whence
\begin{equation}
\mathrm{Re}\braket{\hat a^2}\le\sqrt{N(N+1)} .
\end{equation}
Strong duality holds: the $\hat c$-vacuum at $r=r_\ast$ is a squeezed state of $\hat a$ with $\tanh2r_\ast=-2/\lambda_\ast$, whose mean occupation is
$\sinh^2r_\ast=(\cosh2r_\ast-1)/2=N$, saturating the constraint, and whose
$\braket{\hat a^2}=-\sinh r_\ast\cosh r_\ast=+\sqrt{N(N+1)}$ with the above sign convention. Therefore the bound is tight, and equality holds iff $\ket{\psi}$ is this squeezed vacuum up to a phase (the phase rotation used to reduce $|\braket{\hat a^2}|$ to $\mathrm{Re}\braket{\hat a^2}$ is arbitrary). This proves Eq.~(\ref{eq:momentbound}).

The proof calls for three comments. The dual formulation is what makes the result global: no ansatz on the structure of the optimizing state was made, so the squeezed vacuum beats every competitor, Gaussian or not. The uniqueness of the optimizer follows from the strict concavity of the bound in the relevant interval and from the equality condition in the first step, and it is what upgrades Theorem~3 from ``an optimal state exists'' to ``the optimal state is unique''. The comparison values are instructive: $|\braket{\hat a^2}|=N$ for coherent states, which have maximal single-photon coherence but only $N$-scale two-photon coherence, and $(N-1)/2$ for the two-Fock superpositions $(\ket{N-1}+\ket{N+1})/\sqrt2$, which show that even deliberately chosen non-Gaussian superpositions fall short of the squeezed vacuum by a linear factor. Numerical maximization over truncated Fock spaces ($n\le80$, multi-start constrained optimization) reproduces $\sqrt{N(N+1)}$ to $10^{-6}$ for $N=1,2,3$.

\subsection{\label{sec:bestref}Best reference at fixed energy: double-squeezed vacuum}

The moment bound~(\ref{eq:momentbound}) was stated for the optical mode, but nothing in its proof used which mode the operator belongs to; applied to the mechanical mode it gives $|B|\le\sqrt{M(M+1)}$, with equality iff $\ket{\phi}$ is likewise a squeezed vacuum. Inserting both bounds into Eq.~(\ref{eq:varfixedref}) yields the global optimum among product states at fixed mean occupations $(N,M)$:
\begin{equation}
\label{eq:doublesq}
F_{Q,\mathrm{prod}}^{(\theta),\max}=4\Big[(2M+1)N+M
+2\sqrt{N(N+1)M(M+1)}\Big],
\end{equation}
attained by two phase-matched squeezed vacua, of the kind whose parametric generation in cavity optomechanics was analysed in Ref.~\cite{LiaoLaw2011squeezing}. The result has a clean factorized structure: the baseline $2NM+N+M$ is fixed by the energies, while the gain over it is the geometric mean of the two maximal two-photon coherences, so both modes contribute on equal footing, a symmetric role that will recur in Sec.~\ref{sec:joint}. For $N=M$ this gives $F_Q^{(\theta)}=4E^2+8E$ at mean total energy $E=N+M$: heuristically, Heisenberg scaling $E^2$ plus a variance-induced correction $8E$ (see the discussion in Sec.~\ref{sec:joint}). The correction is not an artifact: squeezed vacua have nonzero energy variance, and it is this variance that the pair-coherence term exploits. The enhancement over the balanced Fock--Fock input, $F_Q^{(\theta)}=2E^2+4E$, is a factor approaching $2$ at large $E$: squeezing buys, over the best number state, the pair-coherence contribution that number states cannot have at all.

\section{\label{sec:mechanical}Fixed optical state: optimal mechanical probe}

Because $\hat n$ is invariant under $\hat d\leftrightarrow \hat b$, every statement of Sec.~\ref{sec:optical} holds verbatim with the modes exchanged. In particular, with the optical mode fixed in the Fock state $\ket{n}$, Eq.~(\ref{eq:theorem1}) applies symmetrically,
\begin{equation}
\label{eq:theorem2}
F_Q^{(\theta)}=4\big[(2n+1)\braket{\hat n_b}+n\big]
\quad\text{(any mechanical state)},
\end{equation}
so the optimal mechanical state is degenerate: any state with maximal mean phonon number, e.g.\ a Fock or thermal state, is optimal, and the photon number $n$ of the reference enhances the per-phonon sensitivity by $2n+1$. This is the cleaner of the two ``fixed-reference'' problems experimentally. The optical reference can be prepared deterministically by photon counting, which is a mature capability, while the mechanical state enters the master formula only through $M=\braket{\hat n_b}$; consequently no mechanical state preparation beyond a known energy is required, and even a thermal state of calibrated mean occupation saturates the bound. In practice this means that the $2n+1$ enhancement can be sought without ground-state cooling of the mechanics, provided the phonon occupation is known and stable over the acquisition time.

With a general fixed optical state instead of a Fock reference, the situation mirrors Sec.~\ref{sec:generalref} exactly: the optimal mechanical probe is a phase-matched squeezed vacuum, whose squeezing axis must be aligned with the complex conjugate of the optical reference's pair coherence $\braket{\hat d^2}$, and whose gain is limited by the same moment bound $|\braket{\hat b^{\dagger2}}|\le\sqrt{M(M+1)}$. The maximal QFI at fixed mean occupations $(N,M)$ is given by Eq.~(\ref{eq:doublesq}) with the roles of the modes exchanged, and the phase-matching condition~(\ref{eq:phasematch}) becomes $\arg\braket{\hat b^2}=-\arg\braket{\hat d^{\dagger2}}$. The same phase-sensitivity warnings apply: a mismatch $\varphi=\pi$ between the two squeezing axes flips the sign of the pair-coherence term and drives the QFI below the number-state baseline, so phase control of the mechanical squeezing matters as much as its magnitude.

The beam-splitter interaction is perfectly symmetric under $\hat d\leftrightarrow \hat b$: unlike the full radiation-pressure coupling, whose linearization treats the two modes differently (the drive displaces only the optical mode), the resulting beam-splitter generator treats them on an equal footing, and so do the optimal-state problems. Every closed-form result of Sec.~\ref{sec:optical} (the degenerate number-reference optimum, the moment bound, the phase-matched squeezed optimum, and the double-squeezed product optimum) therefore has a mechanical-mode twin, and Tables~\ref{tab:table1} and~\ref{tab:table2} can be read in either direction.

\section{\label{sec:joint}Joint optimization I: definite total energy}

\begin{figure}[t]
\includegraphics[width=\columnwidth]{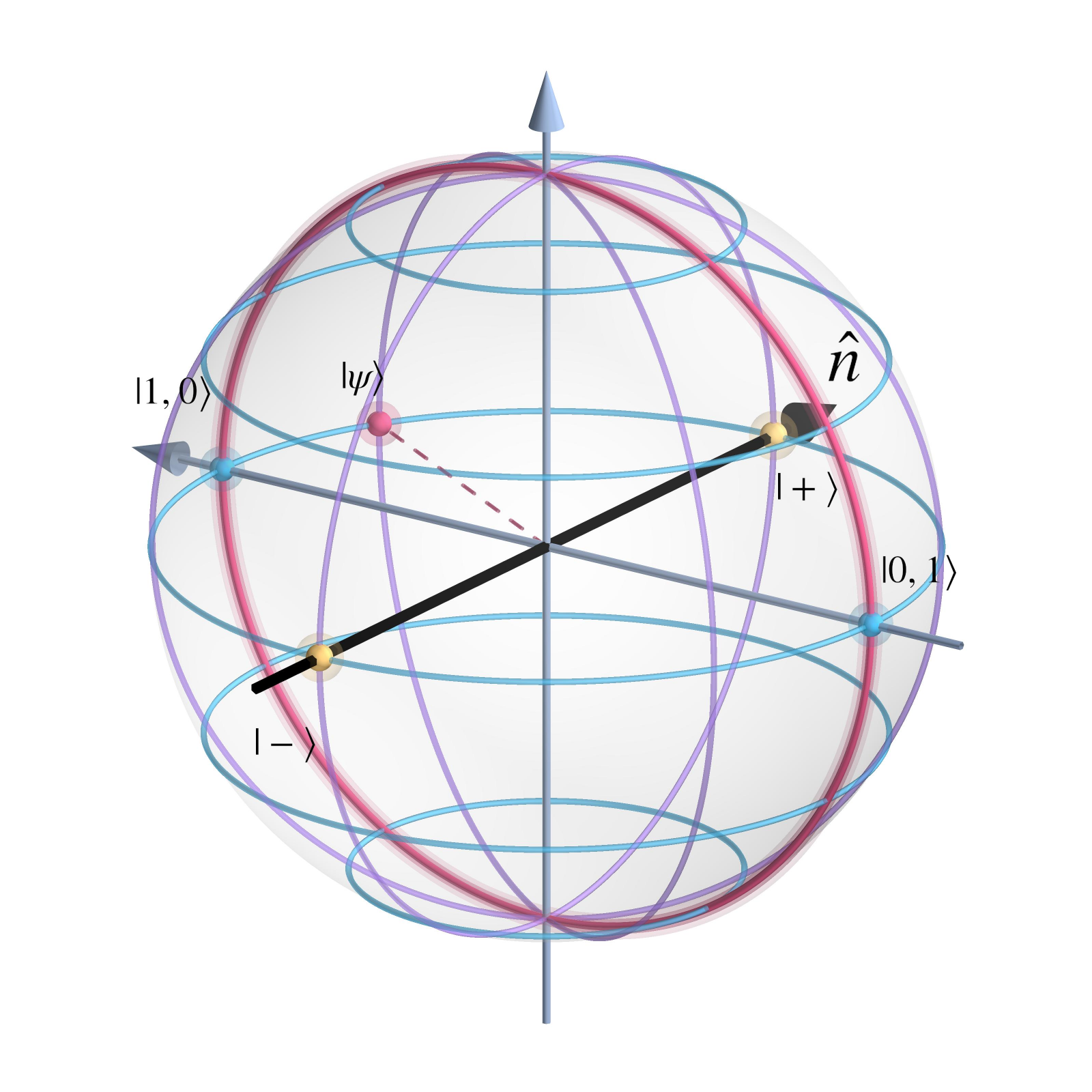}
\caption{\label{fig:bloch}Geometric picture of the definite-energy optimum. Each point of the translucent sphere represents a pure state in the fixed-$E$ $SU(2)$ manifold. The full-diameter black axis is the $\hat n$ direction; its endpoints (gold markers $\ket{+}$ and $\ket{-}$) are the extremal eigenstates, where $\mathrm{Var}(\hat n)=0$. Cyan parallels collect the states of fixed polar angle $\vartheta$, i.e.\ fixed variance $\mathrm{Var}(\hat n)=E^2\sin^2\vartheta$; purple meridians are the constant-phase paths that connect the two poles and parametrize the relative phase $\varphi$ within each parallel. The glowing pink equator ($\vartheta=\pi/2$) is precisely the optimal family~(\ref{eq:optstate}) of equal-weight superpositions of $\ket{\pm}$, on which $\mathrm{Var}(\hat n)=E^2$ and $F_Q^{(\theta)}=4E^2$ are maximal; for $E=1$ its two antipodes are the Fock states $\ket{1,0}$ and $\ket{0,1}$ (cyan markers). The pink marker $\ket{\psi}$ on the $\vartheta=\pi/3$ parallel (dashed radial line) is a generic non-optimal state, whose variance $E^2\sin^2(\pi/3)=3E^2/4$ falls a factor $4/3$ short of the equatorial value.}
\end{figure}

If both modes may be optimized but the total excitation number is fixed, $\hat n_d+\hat n_b=E$, the problem reduces to maximizing $\mathrm{Var}(2\hat J_x)$ within the $(E{+}1)$-dimensional $SU(2)$ manifold. On this manifold $\hat n^2\le E^2\,\mathbb{1}$, hence
\begin{equation}
\mathrm{Var}(\hat n)=\braket{\hat n^2}-\braket{\hat n}^2\le E^2,
\end{equation}
with equality iff the state is supported on the joint eigenspace of $\hat n^2$ with eigenvalue $E^2$ (the two extremal eigenstates $\ket{\pm}\equiv\ket{E/2\ \text{phonons exchanged}}$, eigenvalues $\hat n=\pm E$) and has $\braket{\hat n}=0$, i.e.\ equal weights: writing $\ket{\psi}=c_+\ket{+}+c_-\ket{-}$, equality requires $\braket{\hat n}=E(|c_+|^2-|c_-|^2)=0$, hence $|c_+|=|c_-|$. The bound is sharp. The cruder chain $\mathrm{Var}(\hat n)=4\,\mathrm{Var}(\hat J_x)\le4\braket{\hat J^2}=4j(j+1)=E^2+2E$ with $j=E/2$ is valid but not tight; the operator inequality $\hat n^2\le E^2$ on the fixed-$E$ manifold is what yields the sharp constant. The definite-energy optimum is therefore the family
\begin{equation}
\label{eq:optstate}
\ket{\Psi_{\rm opt}^{(\varphi)}}=\frac{1}{\sqrt2}\left(\ket{+}+e^{i\varphi}\ket{-}\right),
\qquad F_Q^{(\theta)}=4E^2 .
\end{equation}
These optimal states are highly nonclassical two-mode entangled states. Figure~\ref{fig:bloch} makes the geometric content of this result explicit. Since every pure state of the fixed-$E$ manifold can be written as a spin-$j=E/2$ state, the manifold is a Bloch sphere: each point of the sphere is one pure state, the polar axis of the sphere is the $\hat n$ direction, and a state at polar angle $\vartheta$ from the $+\hat n$ pole satisfies $\mathrm{Var}(\hat n)=E^2\sin^2\vartheta$. The two poles (the intersection of the sphere with the full-diameter $\hat n$ axis drawn in the figure) are the extremal eigenstates $\ket{\pm}$ themselves, the only states with $\mathrm{Var}(\hat n)=0$; every state that leaves a pole carries a nonzero variance, growing as $\sin^2\vartheta$ along the parallels (the cyan circles of fixed $\vartheta$ in the figure) and peaking on the equator. The meridians (purple) complete the parametrization: they connect the two poles along paths of fixed relative phase $\varphi$, so the optimal family~(\ref{eq:optstate}), which has $|c_+|=|c_-|=1/\sqrt2$ and phases $e^{i\varphi}$, is the glowing equator of the figure. Two further landmarks are marked for orientation: for $E=1$ the equatorial antipodes $\ket{1,0}$ and $\ket{0,1}$ are the balanced Fock states, i.e.\ the images of the poles under the beam-splitter rotation constructed below, while the off-equator marker $\ket{\psi}$ at $\vartheta=\pi/3$ illustrates a generic non-optimal state whose variance $3E^2/4$ already falls short of the equatorial maximum. A convenient construction identifies $\ket{\pm}$ as the images of the Fock states $\ket{E,0}$ and $\ket{0,E}$ under the rotation $e^{-i(\pi/2)\hat J_y}$, the unitary generated by the interaction Hamiltonian with $\hat J_x\to\hat J_y$, physically the analog of a balanced beam splitter acting on the two-mode NOON state:
\begin{equation}
\ket{\Psi_{\rm opt}}=e^{-i(\pi/2)\hat J_y}\,
\frac{\ket{E,0}+e^{i\phi}\ket{0,E}}{\sqrt2}
\end{equation}
(up to member-dependent phases within the degenerate family; our numerics confirm $\mathrm{Var}(\hat n)=E^2$ and that the rotated NOON states lie in the span of $\{\ket{+},\ket{-}\}$). This mirrors the usual interferometric statement that the maximal phase sensitivity at fixed particle number is obtained by equal superpositions of the extremal generator eigenstates~\cite{Pezze2018,Dorner2009}.

For comparison, product Fock inputs $\ket{n}\ket{m}$ on the same manifold give [Eq.~(\ref{eq:theorem1}) with $N=n$, $m$ fixed]
\begin{equation}
\label{eq:fockpair}
F_Q^{(\theta)}=4\,(2nm+n+m),
\end{equation}
maximized at fixed $E$ by the balanced choice $n=m=E/2$, where $F_Q^{(\theta)}=2E^2+4E$: exactly half the Heisenberg value to leading order. The single-photon state $\ket{1,0}$ ($E=1$) is the special case in which all strategies coincide, $F_Q^{(\theta)}=4$.

\begin{table*}[t]
\caption{\label{tab:table2}Maximal $F_Q^{(\theta)}$ under different energy constraints at total energy $E$. The double-squeezed value $4E^2+8E$ refers to product states at equal mean occupations $N=M=E/2$; it exceeds $4E^2$ because the energy variance is nonzero, cf.\ Sec.~\ref{sec:joint}.}
\begin{ruledtabular}
\scriptsize
\begin{tabular}{llc}
constraint & optimal state & $F_Q^{(\theta),\max}$ \\
\hline
definite energy $E$ &
extremal-superposition family~(\ref{eq:optstate}) &
$4E^2$ \\
product only, $\braket{\hat n_d}=\braket{\hat n_b}=E/2$ &
phase-matched double squeezed &
$4E^2+8E$ \\
definite energy, product only &
balanced Fock--Fock $\ket{E/2,E/2}$ &
$2E^2+4E$ \\
coherent probe, vacuum reference &
coherent state $\ket{\sqrt E}$ &
$4E$ \\
\end{tabular}
\end{ruledtabular}
\end{table*}

\begin{figure}[t]
\includegraphics[width=\columnwidth]{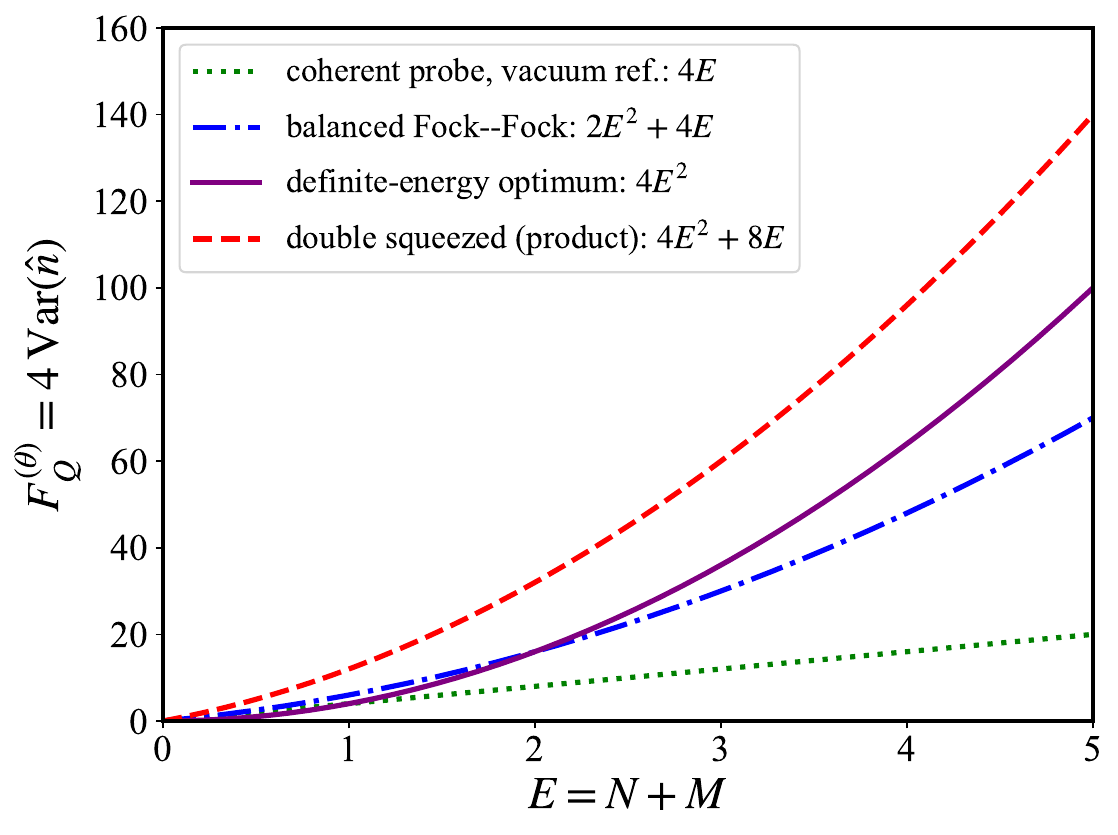}
\caption{\label{fig:scaling}QFI (for angle estimation, $F_Q^{(\theta)}=4\,\mathrm{Var}(\hat n)$) as a function of the total mean excitation number $E$ for the strategies of Table~\ref{tab:table2}. The definite-energy optimum $4E^2$ is the Heisenberg envelope; the phase-matched double-squeezed pair exceeds it at equal mean energy because its energy variance is nonzero.}
\end{figure}

\subsection{A remark on energy constraints}

The precise formulation of the energy constraint matters. If only the mean total energy is bounded, the QFI is unbounded: the states
\begin{equation}
\ket{\psi_\epsilon}=\sqrt{1-\epsilon}\,\ket{0,0}
+\sqrt{\epsilon}\,\ket{+}_{E'},\qquad
\epsilon=\tilde E/E',
\end{equation}
with $\ket{\pm}_{E'}$ the extremal eigenstates on the manifold of total excitation $E'$, have a fixed mean energy $\tilde E$ but $\mathrm{Var}(\hat n)\simeq\epsilon E'^2=\tilde E^2/\epsilon\to\infty$. This ``super-Heisenberg'' divergence from heavy-tailed superpositions is the optomechanical counterpart of the well-known unboundedness of QFIs at fixed mean energy, and is the reason we have formulated the joint optimization at definite energy, where the sharp bound $4E^2$ holds. Within the class of product states the unboundedness is cured by the moment bound~(\ref{eq:momentbound}), giving the finite value~(\ref{eq:doublesq}). Table~\ref{tab:table2} and Fig.~\ref{fig:scaling} summarize the scaling of all strategies.

\section{\label{sec:blue}Joint optimization II: the blue sideband}

On the blue mechanical sideband, $\Delta=+\omega_m$, the rotating-wave approximation retains the counter-rotating (two-mode-squeezing) part of the linearized interaction, Eq.~(\ref{eq:fourterms}),
\begin{equation}
\label{eq:hintblue}
\hat H_{\rm int}^{(\rm b)}=\hbar G\,\hat s,\qquad
\hat s\equiv \hat d\,\hat b+\hat d^\dagger \hat b^\dagger ,
\end{equation}
and the fixed-generator formalism of Sec.~\ref{sec:qfi} applies verbatim: for any pure input, $F_Q^{(\theta)}=4\,\mathrm{Var}(\hat s)$, with the variance again evaluated on the initial state. The generator $\hat s$, however, has a structure fundamentally different from that of $\hat n$, and the optimization problems solved in Secs.~\ref{sec:optical}--\ref{sec:joint} change character. Three structural facts organize the analysis. First, $\hat s$ conserves the excitation difference $\hat n_d-\hat n_b$ rather than the sum, so the Hilbert space decomposes into invariant sectors of fixed particle-number difference, each carrying a positive-discrete-series representation of $su(1,1)$: with $\hat K_0=(\hat n_d+\hat n_b+1)/2$, $\hat s=2\hat K_x$ and $[\hat K_-,\hat K_+]=2\hat K_0$ for $\hat K_\mp=(\hat d\,\hat b,\ \hat d^\dagger\hat b^\dagger)$. Second, the spectrum of $\hat s$ is unbounded from both sides: $\hat s$ possesses no normalizable extremal eigenstates, and the Casimir identity $\hat K_x^2\le\hat K_0^2$ replaces the operator bound $\hat n^2\le E^2$ that drove the beam-splitter analysis of Sec.~\ref{sec:joint}. Third, the interaction is parametrically unstable, since the joint quadratures grow exponentially under $\hat H_{\rm int}^{(\rm b)}$; the following is therefore a short-pulse treatment, valid for $Gt\lesssim1$ and for couplings below the instability threshold of the driven cavity; the QFI itself, being an initial-state quantity, is unaffected by this caveat.

\subsection{Definite initial energy: balanced Fock state}

\begin{figure}[t]
\includegraphics[width=\columnwidth]{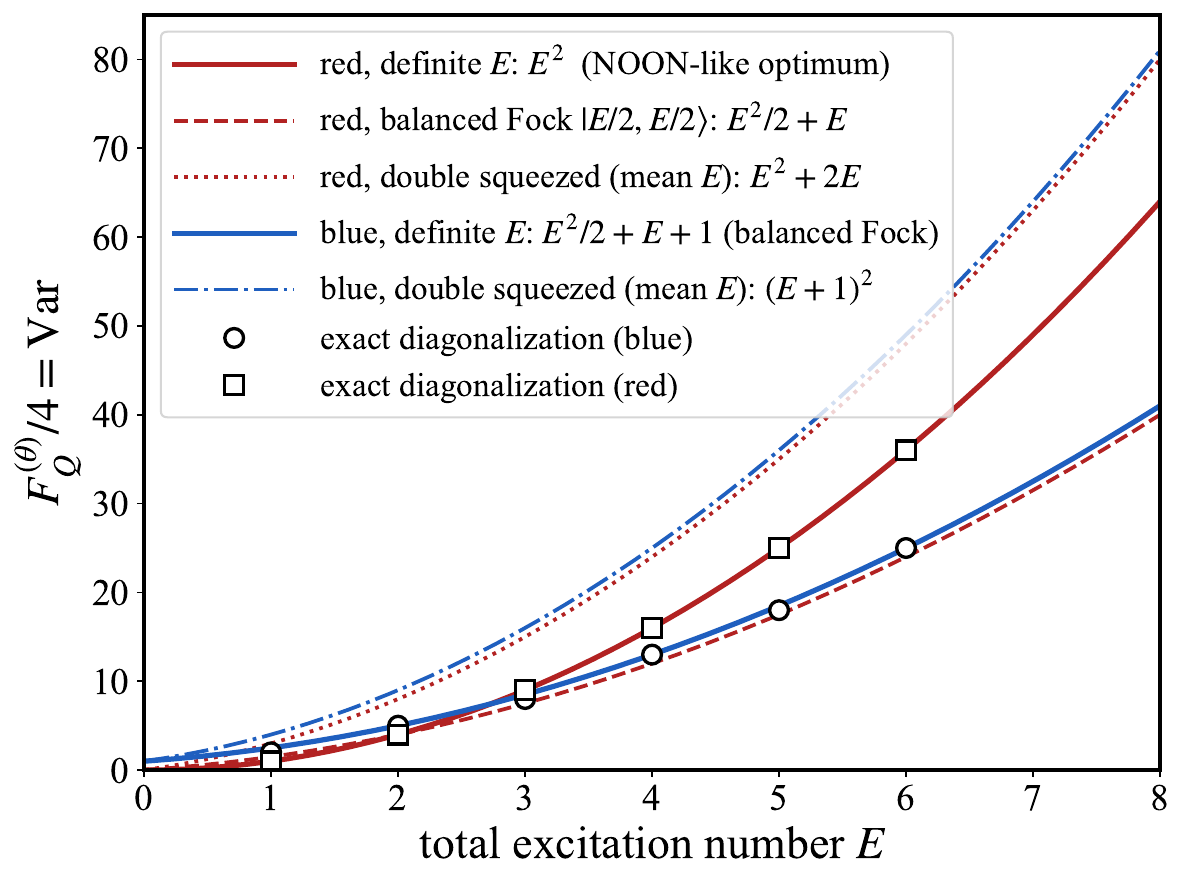}
\caption{\label{fig:blue}Red (beam-splitter) versus blue (two-mode-squeezing) sideband at fixed energy budget $E$. Definite-energy optima: red, $E^2$ from the extremal-superposition family~(\ref{eq:optstate}); blue, $E^2/2+E+1$ from the balanced Fock state~(\ref{eq:blueopt}) (for odd $E$ the exact maximum sits $1/2$ below the even-$E$ envelope). Product-state optima at fixed mean occupations: double-squeezed pairs, $E^2+2E$ (red) versus $(E+1)^2$ (blue) at $N=M=E/2$; balanced Fock, $E^2/2+E$. Circles and squares: exact diagonalization. The blue sideband wins at small $E$; the red sideband wins the Heisenberg asymptote.}
\end{figure}

Although $\hat s$ does not conserve the total energy, the input can of course be prepared in the definite-energy manifold $\{\ket{k,E-k}\}_{k=0}^{E}$. Since $\hat s\ket{\psi}$ then lies entirely in the $E\pm2$ sectors, $\braket{\hat s}=0$ automatically and the variance reduces to $\braket{\hat s^2}$. Squaring the generator,
\begin{equation}
\hat s^2=\hat d^2\hat b^2+\hat d^{\dagger2}\hat b^{\dagger2}
+(\hat n_d+1)(\hat n_b+1)+\hat n_d(\hat n_b+1),
\end{equation}
the pair terms change the energy by $\pm4$ and have vanishing expectation values on $\ket{\psi}$, while the number terms are diagonal on the manifold, giving
\begin{equation}
\label{eq:blueE}
\mathrm{Var}(\hat s)=2\braket{\hat n_d\hat n_b}+E+1
\;\le\;2\max_k\,k(E-k)+E+1 .
\end{equation}
For even $E$ the maximum is attained uniquely by the balanced Fock state $\ket{E/2,E/2}$,
\begin{equation}
\label{eq:blueopt}
F_Q^{(\theta)}=2E^2+4E+4 ,
\end{equation}
while for odd $E$ the maximum $\tfrac12(E^2-1)+E+1$ is attained by any state supported on the two balanced basis vectors $\{\ket{\lfloor E/2\rfloor,\lceil E/2\rceil},\ket{\lceil E/2\rceil,\lfloor E/2\rfloor}\}$. [Both statements are confirmed by exact diagonalization; see the circles in Fig.~\ref{fig:blue}.] Two aspects of the result stand out. The optimal state is a product state: the entangled NOON-type construction of Sec.~\ref{sec:joint} is not merely unnecessary on the blue sideband, it is impossible, because $\hat s$ has no extremal eigenstates to superpose. And the optimal value exceeds the balanced-Fock value $2E^2+4E$ of the red sideband by exactly $+4$: the c-number part $[\hat d\,\hat b,\hat d^\dagger\hat b^\dagger]=\hat n_d+\hat n_b+1$ of the two-mode-squeezing commutator contributes one extra unity to $\braket{\hat s^2}$, a remnant of the pump quantum supplied by the drive at each elementary process. Relative to the definite-energy Heisenberg bound $4E^2$ of the beam-splitter interaction, however, the blue-sideband optimum retains only half of the leading coefficient: the absence of extremal eigenstates costs a factor of two at large $E$ (Fig.~\ref{fig:blue}), while at small $E\lesssim3$ the blue sideband is actually the better channel.

\subsection{Fixed mean occupations: double-squeezed vacuum again}

For product inputs at fixed mean occupations $(N,M)$ the expansion of $\mathrm{Var}(\hat s)$ reads
\begin{equation}
\label{eq:bluemaster}
\mathrm{Var}(\hat s)=2\,\mathrm{Re}\!\left[\braket{\hat d^2}\braket{\hat b^2}\right]
+2NM+N+M+1
-4\,\mathrm{Re}^2\!\left[\braket{\hat d}\braket{\hat b}\right],
\end{equation}
which differs from Eq.~(\ref{eq:master}) only in the pairing $\braket{\hat b^2}$ in place of $\braket{\hat b^{\dagger2}}$ and in the $+1$ contributed by the commutator. The moment bound~(\ref{eq:momentbound}) applies unchanged to each mode, so the global product optimum at fixed $(N,M)$ is again a pair of squeezed vacua, now with the phase-matching condition $\arg\braket{\hat d^2}=-\arg\braket{\hat b^2}$ (squeeze axes aligned with each other rather than mirror-imaged), and
\begin{equation}
\label{eq:bluedouble}
F_{Q,\mathrm{prod}}^{(\theta),\max}
=4\Big[2\sqrt{N(N+1)M(M+1)}+2NM+N+M+1\Big],
\end{equation}
reducing to $4(2N+1)^2$ for $N=M$: exactly the red-sideband product optimum $4E^2+8E$ (at $E=2N$) plus $4$. The parallel with Sec.~\ref{sec:bestref} is complete: on both sidebands the phase-matched double-squeezed vacuum is the optimal product input. The definite-energy optima, by contrast, differ in structure (balanced Fock product state versus NOON-like entanglement) because of the unboundedness of $\hat s$.

Finally, at fixed mean total energy without definiteness, the blue-sideband QFI is unbounded by the same heavy-tailed construction as in Sec.~\ref{sec:joint} (superpose the vacuum with a small-amplitude strongly two-mode-squeezed component), so the definite-energy and fixed-$(N,M)$ formulations above are the meaningful ones, just as on the red sideband.

\section{\label{sec:open}Open-system dynamics: from $t^2$ to $t$}

Up to this point the excitation exchange has been treated as lossless. In reality the optical mode decays at rate $\kappa$ and the mechanical mode couples at rate $\gamma_m$ to a thermal bath of mean occupation $\bar n_{\rm th}$, so that the encoding becomes a noisy quantum channel. In the drive frame and within the linearized regime, the joint state of the two fluctuation modes obeys the Lindblad master equation
\begin{equation}
\label{eq:lindblad}
\dot\rho=-iG[\hat n,\rho]
+\kappa\,\mathcal{D}[\hat d]\rho
+\gamma_m(\bar n_{\rm th}+1)\,\mathcal{D}[\hat b]\rho
+\gamma_m\bar n_{\rm th}\,\mathcal{D}[\hat b^\dagger]\rho ,
\end{equation}
with $\mathcal{D}[\hat o]\rho=\hat o\rho\hat o^\dagger-\tfrac12\{\hat o^\dagger\hat o,\rho\}$ and mechanical decoherence rate $\Gamma_m\equiv\gamma_m(2\bar n_{\rm th}+1)$; the parameter $G=g_0\alpha$ is imprinted exclusively through the Hamiltonian part. Throughout this section the sensing window is the pulsed one of the preceding sections (prepare, sense, read out), with the drive maintaining the coherent amplitude $\alpha$.

\subsection{Short times: the universal quadratic regime and its exact form}

For $t\ll\kappa^{-1}$ the evolution is essentially unitary and all ideal results hold with small corrections,
\begin{equation}
F_Q^{(G)}(t)=4t^2\,\mathrm{Var}(\hat n)\,
\big[1-O\big((\kappa+\Gamma_m)t\big)\big].
\end{equation}
The simplest input can be solved exactly and already displays the full structure. For $\ket{1}_d\ket{0}_b$, the no-jump evolution closes on the pair $\{\ket{1,0},\ket{0,1}\}$: the amplitudes obey the damped two-level problem
\begin{equation}
a(t)=e^{-\kappa t/4}\Big[\cos\Omega t-\tfrac{\kappa}{4\Omega}\sin\Omega t\Big],
\qquad
s(t)=-i\,e^{-\kappa t/4}\,\tfrac{G}{\Omega}\sin\Omega t,
\end{equation}
with $\Omega=\sqrt{G^2-\kappa^2/16}$, and every photon jump leaves the parameter-independent state $\ket{0,0}$. The resulting state is a mixture of the no-jump vector and the vacuum, and to leading order in $G/\kappa$ its QFI for $G$ is
\begin{equation}
\label{eq:openuni}
F_Q^{(G)}(t)=\frac{16}{\kappa^2}\Big(1-e^{-\kappa t/2}\Big)^2
\;\Big[1+O\big((G/\kappa)^2\big)\Big]:
\end{equation}
it reproduces the ideal $4t^2$ at $t\ll\kappa^{-1}$ and then saturates at $16/\kappa^2$ for $t\gg\kappa^{-1}$ rather than decaying. The saturation has a direct physical origin. Once the photon is gone, the memory of $G$ survives in the small probability $p=4G^2/\kappa^2$ that the excitation was exchanged into a phonon before the photon decayed, and a $G$-dependent population $p$ contributes $(\mathrm{d}p/\mathrm{d}G)^2/[p(1-p)]=16/\kappa^2$ no matter how small $p$ is. (With mechanical damping included, the stored phonon also relaxes and the saturation gives way to an eventual decay on the timescale $\Gamma_m^{-1}$.) Exact numerical propagation of Eq.~(\ref{eq:lindblad}) in truncated Fock spaces confirms Eq.~(\ref{eq:openuni}) to $10^{-3}$ relative accuracy at $G/\kappa=10^{-2}$, and shows the same short-time quadratic regime with decoherence-induced bending for every strategy of Secs.~\ref{sec:optical}--\ref{sec:joint}; see Fig.~\ref{fig:open}(a).

\subsection{Long times: renewal at the optimal trial duration, and the $t^2\to t$ transition}

\begin{figure}[t]
\includegraphics[width=\columnwidth]{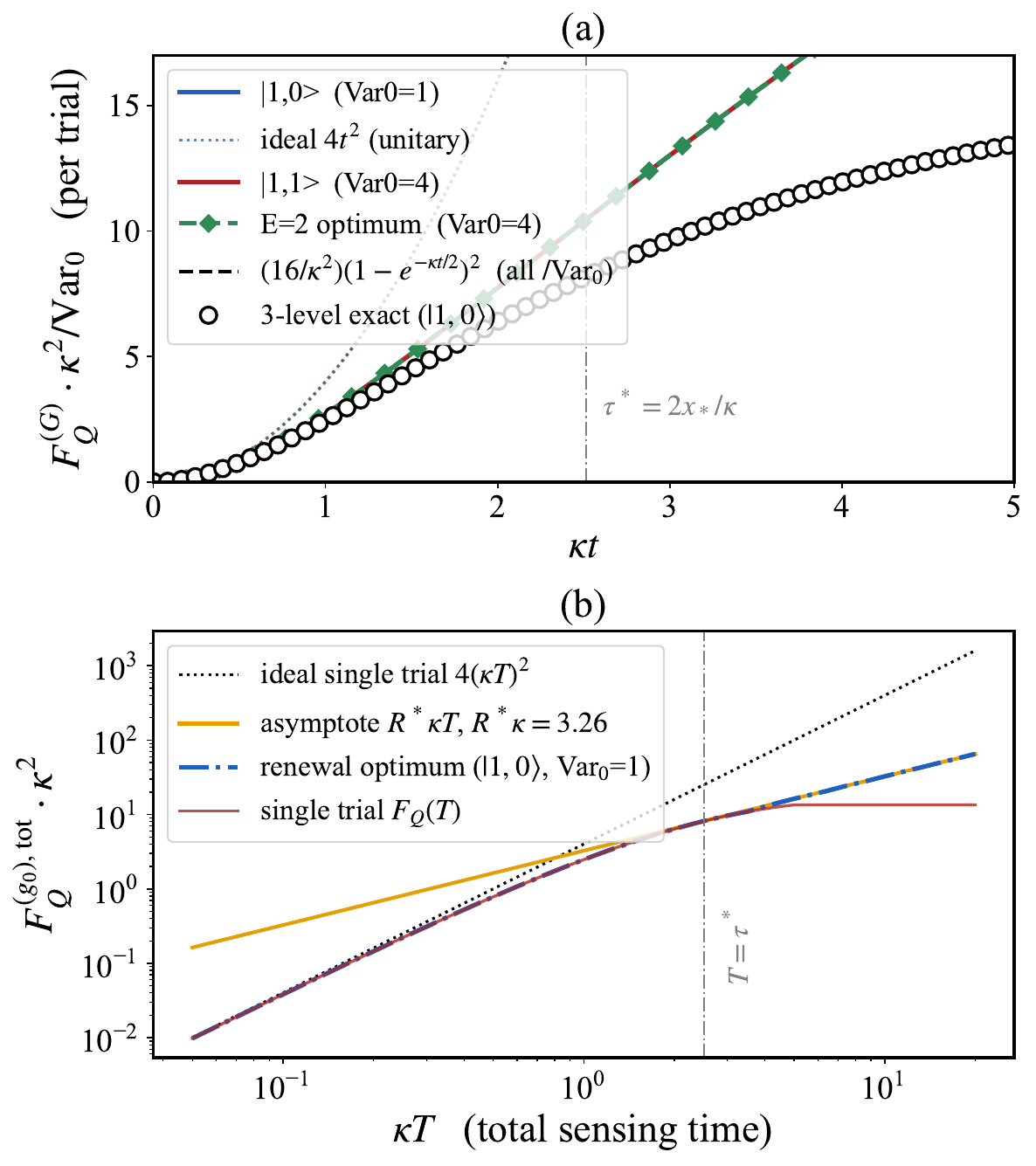}
\caption{\label{fig:open}Open-system dynamics under the master equation~(\ref{eq:lindblad}) with purely optical loss ($\gamma_m=0$), computed exactly in truncated Fock space with $G/\kappa=10^{-2}$. (a)~Per-trial QFI for $G$, in units $\kappa^2/\mathrm{Var}_0$, versus $\kappa t$. All strategies follow the ideal quadratic $4t^2$ (dotted) at $t\ll\kappa^{-1}$ and bend away as loss destroys the excitation exchange. For the single-photon input the exact three-level solution (circles) reproduces the universal curve~(\ref{eq:openuni}) (black dashed), which saturates at $16/\kappa^2$; the $m=1$ reference and the $E=2$ entangled optimum (normalized by their $\mathrm{Var}_0=4$) track their ideal slopes longer but are the first to suffer which-path decoherence. The vertical line marks the renewal-optimal trial duration $\tau^*=2.51\kappa^{-1}$. (b)~Total accumulated QFI versus total sensing time $T$ (log-log) for the $|1,0\rangle$ input: single long trial (red) saturates, while the renewal protocol optimized over $\tau$ (blue dash-dot) crosses over from the ideal $4(\kappa T)^2$ (dotted) to the linear asymptote $R_*\kappa T$ with $R_*\kappa=3.26$ (solid orange); the transition occurs at $T\sim\tau^*$.}
\end{figure}

The pulsed protocol can be repeated. If each independent trial of duration $\tau$ yields $F_Q(\tau)$, the total QFI accumulated in a campaign of duration $T$ is $F_Q^{\rm tot}=\nu F_Q(\tau)$ with $\nu=T/\tau$ repetitions, and the figure of merit is the information rate $R(\tau)=F_Q(\tau)/\tau$. In the ideal unitary case $R=4\,\mathrm{Var}(\hat n)\,\tau$ grows without bound; a single long trial is optimal, which is the quadratic Heisenberg regime. With optical loss, $R(\tau)$ has a finite maximum: for the single-photon input, Eq.~(\ref{eq:openuni}) gives
\begin{equation}
R(\tau)=\frac{16}{\kappa^2\,\tau}\Big(1-e^{-\kappa\tau/2}\Big)^2 ,
\end{equation}
which is maximal at $\tau^*=2x_*/\kappa\simeq2.51\,\kappa^{-1}$, where $x_*$ is the (unique positive) solution of $e^{x}=1+2x$, $x_*\simeq1.2564$. The optimal rate is
\begin{equation}
\label{eq:rate}
R_*=\frac{3.26}{\kappa}\qquad
\big(\ket{1}_d\ket{0}_b,\;G\ll\kappa\big),
\end{equation}
and for $T\gg\tau^*$ the accumulated QFI is linear in the total time, $F_Q^{\rm tot}=R_*T$. This is the advertised transition: the ideal $t^2$ scaling of the loss-free channel crosses over to the standard linear (shot-noise-like) accumulation of information, and the transition point is set by the cavity lifetime, $t_c\sim\kappa^{-1}$ [more precisely, the optimal trial duration $\tau^*=2.51\kappa^{-1}$ marks the end of the quadratic regime and the asymptotes $4t^2$ and $R_*t$ cross at $t=0.82\,\kappa^{-1}$; Fig.~\ref{fig:open}(b)]. The physical content is straightforward: the ideal prefactor $\alpha^2t^2$ in Eq.~(\ref{eq:chain}) is cut off at an effective interaction time of order the cavity lifetime, because it is the photon flux through the cavity (each photon dwelling for a time $\sim\kappa^{-1}$) that limits the rate at which information about $g_0$ can be accumulated.

\subsection{Robustness of the strategies against decoherence}

The strategies of this paper respond very differently to the two baths. Number and thermal references (Theorems~1 and~2) are governed by a master formula involving only the mean occupations, so mechanical damping merely reduces $M(t)=Me^{-\gamma_m t}$ (negligible over the optimal trial time $\tau^*\sim\kappa^{-1}$ whenever $\gamma_m\ll\kappa$), and a thermal mechanical state of calibrated occupation is as robust as a Fock one; these strategies are limited by optical loss alone. Pair coherences are far more fragile: the mechanical squeezing $\braket{\hat b^{\dagger2}}$ entering Eqs.~(\ref{eq:phasematch}) and~(\ref{eq:doublesq}) is not a population but a coherence, and it decays at the mechanical decoherence rate $\Gamma_m=\gamma_m(2\bar n_{\rm th}+1)$, which in optical experiments exceeds $\gamma_m$ by the thermal factor $2\bar n_{\rm th}+1$ (for the parameters of Ref.~\cite{Verhagen2012}, $\Gamma_m/2\pi=5.6\ \mathrm{MHz}\simeq0.8\,\kappa$); pair-coherence gains are therefore suppressed by $e^{-\Gamma_m\tau}$ and survive only if $\Gamma_m\lesssim\kappa$. The same fragility afflicts entangled states of definite energy: a lost photon reveals which component of a NOON-type superposition it left, so the generator-eigenstate coherence decays at a rate proportional to the optical excitation number $E$, and the Heisenberg window shrinks as $t\ll(E\kappa)^{-1}$, which is the usual mechanism by which lossy interferometry forfeits the ideal $4E^2$ envelope and reverts to linear scaling with an $E$-independent slope~\cite{Escher2011,Demkowicz2012}. The practical conclusion, quantified in the next section, is that at the few-quanta level accessible today, the number-reference and squeezed-probe strategies are the robust ones, while the entangled optima demand both high cooperativity and low mechanical decoherence.

\section{\label{sec:crb}Estimating $g_0$: precision, scaling, and outlook}

Collecting the results, the ultimate precision of estimating the single-photon coupling after $\nu$ independent repetitions is
\begin{equation}
\Delta g_0\ \ge\ \frac{1}{\alpha t\,\sqrt{\nu\,F_Q^{(\theta)}}}\,,
\end{equation}
with $F_Q^{(\theta)}$ given by the entries of Tables~\ref{tab:table1} and~\ref{tab:table2}. Three practical levers emerge, in increasing order of power: a photon/phonon-number enhancement $2m+1$ or $2n+1$ from a number-state reference, which costs no probe energy; squeezing on both sides, which adds the pair-coherence term $2\sqrt{N(N+1)M(M+1)}$; and entanglement across the excitation manifolds, which reaches $4E^2$ at definite energy. All three are multiplicative and compatible.

\subsection{\label{sec:realistic}A realistic parameter case study}

To answer the natural objection that the preceding treatment is idealized, we combine the open-system analysis with the parameters of the state-of-the-art experiment of Verhagen et al.~\cite{Verhagen2012}, which achieved quantum-coherent optomechanical coupling in the resolved-sideband regime: $g_0/2\pi=3.4$ kHz (at 780 nm), $\kappa/2\pi=7.1$ MHz (time-domain strong-coupling run), $\omega_m/2\pi=78.2$ MHz, $\gamma_m/2\pi=3.6$ kHz, mechanical decoherence rate $\Gamma_m/2\pi=5.6$ MHz (i.e.\ $\bar n_{\rm th}\simeq780$), radiation-pressure cooling of the mechanics to a mean occupation $\bar n_f=1.7\pm0.1$ phonons, and a demonstrated intracavity photon number $n_c=|\alpha|^2=3\times10^5$. The linearized coupling is then $G/2\pi=g_0\sqrt{n_c}/2\pi=1.86$ MHz, i.e.\ $G/\kappa=0.26$: comfortably inside the weak-coupling (linearized) regime assumed throughout, and reached in the laboratory.

\begin{figure}[t]
\includegraphics[width=\columnwidth]{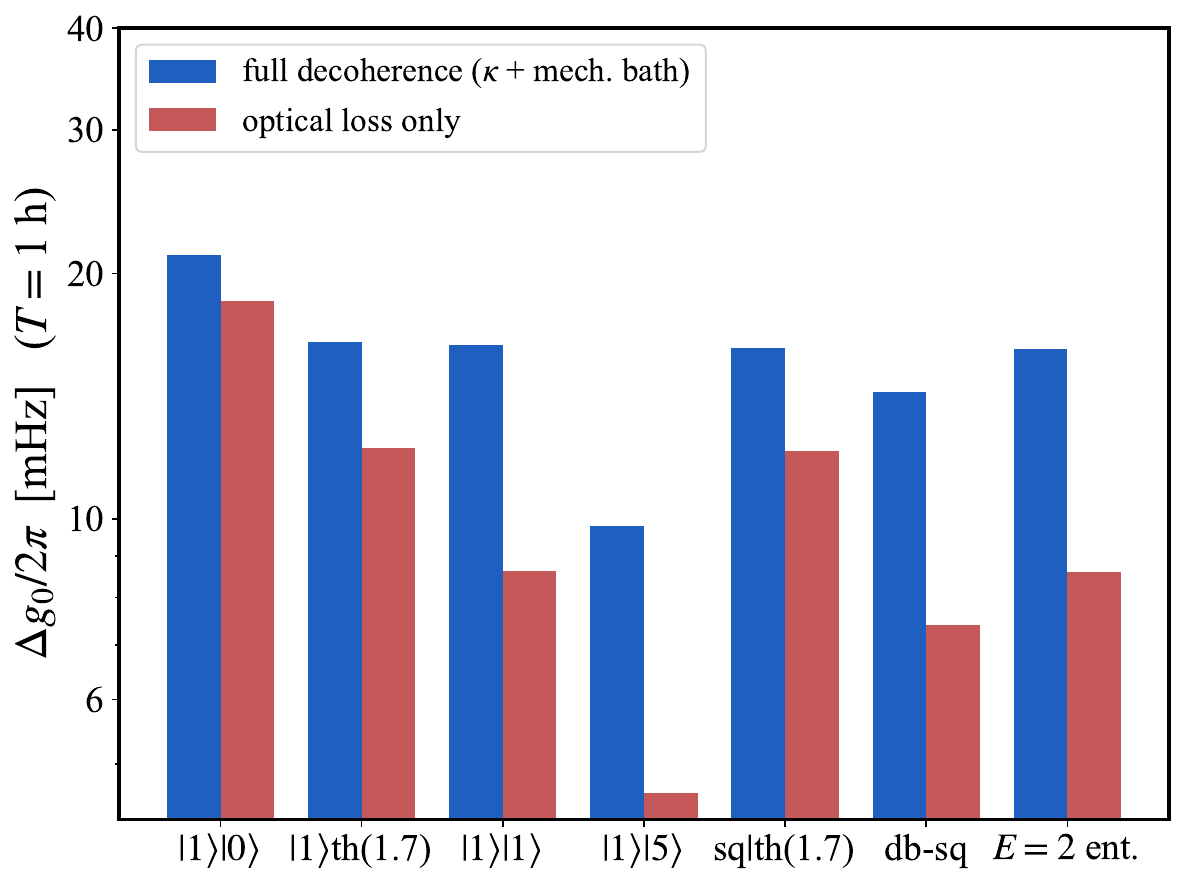}
\caption{\label{fig:realistic}Realistic $g_0$ precision with the Verhagen et al. parameters of Table~\ref{tab:realistic}: $\Delta g_0/2\pi$ after one hour of renewal-optimized pulsed sensing, with the full decoherence of Eq.~(\ref{eq:lindblad}) (blue) and with optical loss only (red). Mechanical decoherence penalizes most strongly the strategies that rely on mechanical coherences (Fock $m=5$, double-squeezed), while the vacuum-reference protocol is nearly unaffected; a squeezed probe adds nothing against a thermal reference, and the $E=2$ entangled state holds no advantage at this few-quanta level.}
\end{figure}

The pulsed protocol is evaluated exactly: the master equation~(\ref{eq:lindblad}) with these parameters is propagated in truncated Fock space for each initial strategy of Tables~\ref{tab:table1} and~\ref{tab:table2}, the per-trial QFI is obtained from the SLD formula, the trial duration is optimized, $\tau^*=\arg\max_\tau F_Q^{(G)}(\tau)/\tau$, and the per-hour precision is $\Delta g_0=[n_c\,R(\tau^*)\,T]^{-1/2}$ with $T=1$ h. The factor $n_c=|\alpha|^2$ converts the QFI of $G$ into that of $g_0$ via Eq.~(\ref{eq:chain}). State preparation and readout are assumed ideal; a total detection efficiency $\eta$ rescales the rates of Sec.~\ref{sec:open} by $\eta$. Table~\ref{tab:realistic} and Fig.~\ref{fig:realistic} collect the results.

\begin{table}[t]
\caption{\label{tab:realistic}Realistic per-hour precision of $g_0$ estimation with the parameters of Ref.~\cite{Verhagen2012} ($g_0/2\pi=3.4$ kHz, $\kappa/2\pi=7.1$ MHz, $\Gamma_m/2\pi=5.6$ MHz, $n_c=3\times10^5$, $T=1$ h, ideal preparation and readout). $\mathrm{Var}_0$ is the initial-state variance of the ideal theory; $\tau^*$ is the renewal-optimal trial duration; the last two columns give $\Delta g_0/2\pi$ with the full decoherence of Eq.~(\ref{eq:lindblad}) and with optical loss only.}
\begin{ruledtabular}
\begin{tabular}{lcccc}
strategy & $\mathrm{Var}_0$ & $\tau^*$ [ns] & \multicolumn{2}{c}{$\Delta g_0/2\pi$ [mHz]} \\
 & & & full & opt.\ only \\
\hline
$\ket{1}_d\ket{0}_b$ \ (state of the art) & 1 & 150 & 21.1 & 18.5 \\
$\ket{1}_d\otimes$thermal$(1.7)$ & 6.1 & 101 & 16.5 & 12.2 \\
$\ket{1}_d\ket{1}_b$ & 4 & 13 & 16.3 & 8.6 \\
$\ket{1}_d\ket{5}_b$ & 16 & 67 & 9.8 & 4.6 \\
squeezed$_{N=1}\otimes$thermal$(1.7)$ & 6.1 & 94 & 16.2 & 12.1 \\
double squeezed $(1.7,1.7)$ & 18.4 & 16 & 14.3 & 7.4 \\
$E=2$ entangled optimum & 4 & 13 & 16.2 & 8.6 \\
\end{tabular}
\end{ruledtabular}
\end{table}

Four lessons follow. First, the precisions are experimentally meaningful: the state-of-the-art single-photon protocol reaches $\Delta g_0/g_0\simeq6\times10^{-6}$ per hour, and all strategies land in the $10^{-6}$--$10^{-5}$ band: absolute precisions of $10$--$20$ mHz on a coupling whose scale is $3.4$ kHz, well below any linewidth of the system. Second, the ideal hierarchy of Table~\ref{tab:table2} is largely preserved but compressed by decoherence: cooling to the experimentally achieved $\bar n_f=1.7$ already improves on the vacuum reference by $\sim$22\%, a mechanical Fock reference with $m=5$ phonons would gain a factor $\sim$2, and the double-squeezed pair retains a $\sim$13\% advantage over the cooled thermal reference even after mechanical decoherence. Third, the pair-coherence structure of the master formula is visible in the data: a squeezed probe is useless against a thermal reference (with $\braket{\hat b^2}=0$ the pair term vanishes and any zero-mean probe of equal energy performs identically; the two corresponding bars coincide), so probe squeezing pays only in combination with a squeezed or phase-coherent reference. Fourth, mechanical decoherence penalizes above all the strategies that rely on mechanical coherences (Fock $m=5$, double-squeezed), roughly doubling their $\Delta g_0$ relative to the optical-loss-only value, while the vacuum-reference protocol pays only $\sim$14\%; and the entangled $E=2$ optimum offers no advantage at this few-quanta level, just as the which-path argument of Sec.~\ref{sec:open} predicts. Since $\Delta g_0\propto\sqrt{\kappa/n_c}$, operating at the intracavity photon numbers of the strong-coupling time-domain run ($n_c\simeq2.8\times10^6$) would bring the best strategies to $\Delta g_0/2\pi\simeq5$ mHz per hour.

Two of the three idealizations of our treatment have now been lifted: Sec.~\ref{sec:blue} resolved the counter-rotating interaction, and Sec.~\ref{sec:open} quantified the effect of cavity loss and mechanical decoherence, including the $t^2\to t$ crossover of the accumulated precision. The remaining idealizations define natural next steps. Readout and state-preparation inefficiencies enter as additional loss channels and further rescale the rates of Sec.~\ref{sec:open}; the pulsed schemes of Ref.~\cite{Vanner2011} provide a route to preparing the required squeezed or number-state references of the mechanical mode; and the strong quantum-coherent coupling demonstrated in Ref.~\cite{Verhagen2012} shows that interaction times of order the cavity lifetime (the regime in which the optimal trial duration $\tau^*$ of Sec.~\ref{sec:open} sits) are experimentally accessible.

\section{\label{sec:conclusion}Conclusion}

We have determined, in closed form, the optimal initial states for estimating the single-photon optomechanical coupling strength $g_0$ in the linearized beam-splitter regime, for pure inputs. The problem reduces to maximizing the variance of the excitation-exchange operator on the initial state. With one mode fixed, the optimal state of the other mode is degenerate over all states of a given energy if the reference is a number state, with a per-particle sensitivity enhancement that grows linearly with the reference phonon number, while for any other reference it is a phase-matched squeezed vacuum, an exact global statement that follows from a new moment bound on the two-photon coherence at fixed mean occupation. With both modes free at definite total energy, equal superpositions of extremal generator eigenstates (beam-splitter-transformed NOON states) saturate a Heisenberg bound that is quadratic in the total energy, while balanced product Fock states reach a smaller value with only a linear energy enhancement. All results were verified by exact diagonalization in truncated Fock spaces, with agreement to numerical precision. The single-photon input frequently used in optomechanical state-exchange schemes is recovered as the simplest saturating case. Three extensions were worked out explicitly. On the blue sideband the two-mode-squeezing generator has an unbounded spectrum, the NOON-type optimum is impossible, and the balanced product Fock state $\ket{E/2,E/2}$ is exactly optimal at definite initial energy, $F_Q=2E^2+4E+4$; at fixed mean occupations the double-squeezed vacuum remains optimal, as on the red sideband. Under Lindblad loss the ideal $t^2$ scaling of the per-trial QFI gives way, through the exact universal curve $F_Q^{(G)}=(16/\kappa^2)(1-e^{-\kappa t/2})^2$ for the single-photon input, to linear-in-$t$ accumulation of information at the renewal-optimal trial duration $\tau^*\simeq2.5\,\kappa^{-1}$ and rate $R_*=3.26/\kappa$; number and thermal references survive decoherence, pair coherences are suppressed by the mechanical decoherence rate, and entangled states suffer which-path decoherence proportional to their energy. Finally, a case study with the parameters of the Verhagen et al.\ experiment shows that relative precisions of order $10^{-6}$--$10^{-5}$ per hour are within reach, and identifies the practically optimal strategies at the few-quanta level. Further extensions (full detection inefficiencies, continuous-monitoring readout, and the unstable strong-coupling regime of the blue sideband) define the natural continuation of this work.

\begin{acknowledgments}
This work was supported by NSFC (Grant No.~12205092), the Hunan Provincial Natural Science Foundation of China (Grant Nos.~2026JJ50100 and 2026JJ50513) and the Scientific Research Fund of Hunan Provincial Education Department (Grant No.~24B0670).
\end{acknowledgments}

\bibliography{refs}

@article{Aspelmeyer2014,
  author  = {Aspelmeyer, Markus and Kippenberg, Tobias J. and Marquardt, Florian},
  title   = {Cavity optomechanics},
  journal = {Reviews of Modern Physics},
  volume  = {86},
  pages   = {1391--1452},
  year    = {2014},
  doi     = {10.1103/RevModPhys.86.1391},
}

@article{Verhagen2012,
  author  = {Verhagen, Ewold and Del{\'e}glise, S{\'e}bastien and Weis, Stefan and Schliesser, Albert and Kippenberg, Tobias J.},
  title   = {Quantum-coherent coupling of a mechanical oscillator to an optical cavity mode},
  journal = {Nature},
  volume  = {482},
  pages   = {63--67},
  year    = {2012},
  doi     = {10.1038/nature10787},
}

@article{Chan2011,
  author  = {Chan, Jasper and Mayer Alegre, Thiago P. and Safavi-Naeini, Amir H. and Hill, Jeff T. and Krause, Alex and Gr{\"o}blacher, Simon and Aspelmeyer, Markus and Painter, Oskar},
  title   = {Laser cooling of a nanomechanical oscillator into its quantum ground state},
  journal = {Nature},
  volume  = {478},
  pages   = {89--93},
  year    = {2011},
  doi     = {10.1038/nature10461},
}

@article{Teufel2011,
  author  = {Teufel, John D. and Donner, Tobias and Li, Dale and Harlow, Jay W. and Allman, Michael S. and Cicak, Katarina and Sirois, Aaron J. and Whittaker, John D. and Lehnert, Konrad W. and Simmonds, Raymond W.},
  title   = {Sideband cooling of micromechanical motion to the quantum ground state},
  journal = {Nature},
  volume  = {475},
  pages   = {359--363},
  year    = {2011},
  doi     = {10.1038/nature10261},
}

@article{Hill2012,
  author  = {Hill, Jeff T. and Safavi-Naeini, Amir H. and Chan, Jasper and Painter, Oskar},
  title   = {Coherent optical wavelength conversion via cavity optomechanics},
  journal = {Nature Communications},
  volume  = {3},
  pages   = {507},
  year    = {2012},
  doi     = {10.1038/ncomms2201},
}

@article{Andrews2014,
  author  = {Andrews, R. W. and Peterson, R. W. and Purdy, T. P. and Cicak, K. and Simmonds, R. W. and Regal, C. A. and Lehnert, K. W.},
  title   = {Bidirectional and efficient conversion between microwave and optical light},
  journal = {Nature Physics},
  volume  = {10},
  pages   = {321--326},
  year    = {2014},
  doi     = {10.1038/nphys2911},
}

@book{Helstrom1976,
  author    = {Helstrom, Carl W.},
  title     = {Quantum Detection and Estimation Theory},
  publisher = {Academic Press},
  address   = {New York},
  year      = {1976}
}

@article{Braunstein1994,
  author  = {Braunstein, Samuel L. and Caves, Carlton M.},
  title   = {Statistical distance and the geometry of quantum states},
  journal = {Physical Review Letters},
  volume  = {72},
  pages   = {3439--3443},
  year    = {1994},
  doi     = {10.1103/PhysRevLett.72.3439},
}

@article{Giovannetti2011,
  author  = {Giovannetti, Vittorio and Lloyd, Seth and Maccone, Lorenzo},
  title   = {Advances in quantum metrology},
  journal = {Nature Photonics},
  volume  = {5},
  pages   = {222--229},
  year    = {2011},
  doi     = {10.1038/nphoton.2011.35},
}

@article{Paris2009,
  author  = {Paris, Matteo G. A.},
  title   = {Quantum estimation for quantum technology},
  journal = {International Journal of Quantum Information},
  volume  = {7},
  pages   = {125--137},
  year    = {2009},
  doi     = {10.1142/S0219749909004839},
}

@article{Degen2017,
  author  = {Degen, C. L. and Reinhard, F. and Cappellaro, P.},
  title   = {Quantum sensing},
  journal = {Reviews of Modern Physics},
  volume  = {89},
  pages   = {035002},
  year    = {2017},
  doi     = {10.1103/RevModPhys.89.035002},
}

@article{Caves1981,
  author  = {Caves, Carlton M.},
  title   = {Quantum-mechanical noise in an interferometer},
  journal = {Physical Review D},
  volume  = {23},
  pages   = {1693--1708},
  year    = {1981},
  doi     = {10.1103/PhysRevD.23.1693},
}

@article{Holland1994,
  author  = {Holland, Murray J. and Burnett, Keith},
  title   = {Interferometric detection of optical phase shifts at the Heisenberg limit},
  journal = {Physical Review Letters},
  volume  = {71},
  pages   = {1355--1358},
  year    = {1993},
  doi     = {10.1103/PhysRevLett.71.1355},
}

@article{Boto2000,
  author  = {Boto, A. N. and Kok, P. and Abrams, D. S. and Braunstein, S. L. and Williams, C. P. and Dowling, J. P.},
  title   = {Quantum interferometric optical lithography: exploiting entanglement to beat the diffraction limit},
  journal = {Physical Review Letters},
  volume  = {85},
  pages   = {2733--2736},
  year    = {2000},
  doi     = {10.1103/PhysRevLett.85.2733},
}

@article{Pezze2018,
  author  = {Pezz{\`e}, Luca and Smerzi, Augusto and Oberthaler, Markus K. and Schmied, Roman and Treutlein, Philipp},
  title   = {Quantum metrology with nonclassical states of atomic ensembles},
  journal = {Reviews of Modern Physics},
  volume  = {90},
  pages   = {035005},
  year    = {2018},
  doi     = {10.1103/RevModPhys.90.035005},
}

@article{Pezze2009,
  author  = {Pezz{\`e}, Luca and Smerzi, Augusto},
  title   = {Entanglement, nonlinear dynamics, and the Heisenberg limit},
  journal = {Physical Review Letters},
  volume  = {102},
  pages   = {100401},
  year    = {2009},
  doi     = {10.1103/PhysRevLett.102.100401},
}

@article{Dorner2009,
  author  = {Dorner, Ulrich and Demkowicz-Dobrzanski, Rafal and Smith, Brian J. and Lundeen, Jeff S. and Wasilewski, Wojciech and Banaszek, Konrad and Walmsley, Ian A.},
  title   = {Optimal quantum phase estimation},
  journal = {Physical Review Letters},
  volume  = {102},
  pages   = {040403},
  year    = {2009},
  doi     = {10.1103/PhysRevLett.102.040403},
}

@article{Escher2011,
  author  = {Escher, B. M. and de Matos Filho, R. L. and Davidovich, L.},
  title   = {General framework for estimating the ultimate precision limit in noisy quantum-enhanced metrology},
  journal = {Nature Physics},
  volume  = {7},
  pages   = {406--411},
  year    = {2011},
  doi     = {10.1038/nphys1958},
}

@article{Demkowicz2012,
  author  = {Demkowicz-Dobrza{\'n}ski, Rafa{\l} and Ko{\l}ody{\'n}ski, Jan and Gu{\c{t}}{\u{a}}, M{\u{a}}d{\u{a}}lin},
  title   = {The elusive Heisenberg limit in quantum-enhanced metrology},
  journal = {Nature Communications},
  volume  = {3},
  pages   = {1063},
  year    = {2012},
  doi     = {10.1038/ncomms2067},
}

@article{Vitali2007,
  author  = {Vitali, David and Gigan, Silvia and Ferreira, Ana and B{\"o}hm, H. R. and Tombesi, Paolo and Guerreiro, A. and Th{\'e}venaz, V. and Aspelmeyer, Markus and Zeilinger, Anton},
  title   = {Optomechanical entanglement between a movable mirror and a cavity field},
  journal = {Physical Review Letters},
  volume  = {98},
  pages   = {030405},
  year    = {2007},
  doi     = {10.1103/PhysRevLett.98.030405},
}

@article{Genes2008,
  author  = {Genes, Claudiu and Vitali, David and Tombesi, Paolo},
  title   = {Simultaneous cooling and entanglement of mechanical modes of a micromirror in an optical cavity},
  journal = {Physical Review A},
  volume  = {77},
  pages   = {050307(R)},
  year    = {2008},
  doi     = {10.1088/1367-2630/10/9/095009},
}

@article{Ou1992,
  author  = {Ou, Z. Y. and Pereira, S. F. and Kimble, H. J. and Peng, K. C.},
  title   = {Realization of the {E}instein-{P}odolsky-{R}osen paradox for continuous variables},
  journal = {Physical Review Letters},
  volume  = {68},
  pages   = {3663--3666},
  year    = {1992},
  doi     = {10.1103/PhysRevLett.68.3663},
}

@article{Monras2006,
  author  = {Monras, Alex},
  title   = {Optimal phase measurements with pure {G}aussian states},
  journal = {Physical Review A},
  volume  = {73},
  pages   = {033821},
  year    = {2006},
  doi     = {10.1103/PhysRevA.73.033821},
}

@article{Vanner2011,
  author  = {Vanner, M. R. and Pikovski, I. and Cole, G. D. and Kim, M. S. and Brukner, {\v{C}}. and Hammerer, K. and Milburn, G. J. and Aspelmeyer, M.},
  title   = {Pulsed quantum optomechanics},
  journal = {Proceedings of the National Academy of Sciences},
  volume  = {108},
  pages   = {16182--16187},
  year    = {2011},
  doi     = {10.1073/pnas.1105098108},
}

@article{Aasi2013,
  author  = {Aasi, J. and others},
  title   = {Enhanced sensitivity of the {LIGO} gravitational wave detector via squeezed vacuum},
  journal = {Nature Photonics},
  volume  = {7},
  pages   = {613--619},
  year    = {2013},
  doi     = {10.1038/nphoton.2013.177},
}

@article{Gigan2006,
  author  = {Gigan, Sylvain and B{\"o}hm, Hannes R. and Paternostro, Mauro and Blaser, Fr{\'e}d{\'e}ric and Langer, Gregor and Hertzberg, Johannes B. and Schwab, Konstantin C. and B{\"a}uerle, Daniel and Aspelmeyer, Markus and Zeilinger, Anton},
  title   = {Self-cooling of a micro-mirror by radiation pressure},
  journal = {Nature},
  volume  = {444},
  pages   = {67--70},
  year    = {2006},
  doi     = {10.1038/nature05273},
}

@article{Arcizet2006,
  author  = {Arcizet, Olivier and Cohadon, Pierre-Fran{\c{c}}ois and Briant, Thomas and Pinard, Michel and Heidmann, Antoine},
  title   = {Radiation-pressure cooling and optomechanical instability of a micromirror},
  journal = {Nature},
  volume  = {444},
  pages   = {71--74},
  year    = {2006},
  doi     = {10.1038/nature05244},
}

@article{WilsonRae2007,
  author  = {Wilson-Rae, I. and Nooshi, N. and Zwerger, W. and Kippenberg, T. J.},
  title   = {Theory of ground state cooling of a mechanical oscillator using dynamic backaction},
  journal = {Physical Review Letters},
  volume  = {99},
  pages   = {093901},
  year    = {2007},
  doi     = {10.1103/PhysRevLett.99.093901},
}

@article{Marquardt2007,
  author  = {Marquardt, Florian and Chen, Joe P. and Clerk, A. A. and Girvin, S. M.},
  title   = {Quantum theory of cavity-assisted sideband cooling of mechanical motion},
  journal = {Physical Review Letters},
  volume  = {99},
  pages   = {093902},
  year    = {2007},
  doi     = {10.1103/PhysRevLett.99.093902},
}

@article{Schliesser2008,
  author  = {Schliesser, Albert and Rivi{\`e}re, R{\'e}mi and Anetsberger, Georg and Arcizet, Olivier and Kippenberg, Tobias J.},
  title   = {Resolved-sideband cooling of a micromechanical oscillator},
  journal = {Nature Physics},
  volume  = {4},
  pages   = {415--419},
  year    = {2008},
  doi     = {10.1038/nphys939},
}

@article{Thompson2008,
  author  = {Thompson, Jack D. and Zwickl, Bradley M. and Jayich, Audra M. and Marquardt, Florian and Girvin, S. M. and Harris, J. G. E.},
  title   = {Strong dispersive coupling of a high-finesse cavity to a micromechanical membrane},
  journal = {Nature},
  volume  = {452},
  pages   = {72--75},
  year    = {2008},
  doi     = {10.1038/nature06715},
}

@article{Kippenberg2008,
  author  = {Kippenberg, Tobias J. and Vahala, Kerry J.},
  title   = {Cavity optomechanics: back-action at the mesoscale},
  journal = {Science},
  volume  = {321},
  pages   = {1172--1175},
  year    = {2008},
  doi     = {10.1126/science.1156032},
}

@article{Poot2012,
  author  = {Poot, Menno and van der Zant, Herre S. J.},
  title   = {Mechanical systems in the quantum regime},
  journal = {Physics Reports},
  volume  = {511},
  pages   = {273--335},
  year    = {2012},
  doi     = {10.1016/j.physrep.2011.12.004},
}

@article{Teufel2011b,
  author  = {Teufel, John D. and Li, Dale and Allman, Michael S. and Cicak, Katarina and Sirois, Aaron J. and Whittaker, John D. and Simmonds, Raymond W.},
  title   = {Circuit cavity electromechanics in the strong-coupling regime},
  journal = {Nature},
  volume  = {471},
  pages   = {204--208},
  year    = {2011},
  doi     = {10.1038/nature09898},
}

@article{Palomaki2013,
  author  = {Palomaki, T. A. and Teufel, J. D. and Simmonds, R. W. and Lehnert, K. W.},
  title   = {Entangling mechanical motion with microwave fields},
  journal = {Science},
  volume  = {342},
  pages   = {710--713},
  year    = {2013},
  doi     = {10.1126/science.1244563},
}

@article{Bochmann2013,
  author  = {Bochmann, Joerg and Vainsencher, Amit and Awschalom, David D. and Cleland, Andrew N.},
  title   = {Nanomechanical coupling between microwave and optical photons},
  journal = {Nature Physics},
  volume  = {9},
  pages   = {712--716},
  year    = {2013},
  doi     = {10.1038/nphys2748},
}

@article{Bagci2014,
  author  = {Bagci, T. and Simonsen, A. and Schmid, S. and Villanueva, L. G. and Zeuthen, E. and Appel, J. and Taylor, J. M. and S{\o}rensen, A. and Schliesser, K. U. S. and Andersen, P. E. and Kippenberg, T. J. and Schliesser, A.},
  title   = {Optical detection of radio waves through a nanomechanical transducer},
  journal = {Nature},
  volume  = {507},
  pages   = {81--85},
  year    = {2014},
  doi     = {10.1038/nature13029},
}

@article{Riedinger2018,
  author  = {Riedinger, Ralf and Wallucks, Andreas and Marinković, Igor and Löschnauer, Clemens and Aspelmeyer, Markus and Hong, Sungkun and Gröblacher, Simon},
  title   = {Remote quantum entanglement between two micromechanical oscillators},
  journal = {Nature},
  volume  = {556},
  pages   = {473--477},
  year    = {2018},
  doi     = {10.1038/s41586-018-0036-z},
}

@article{OckeloenKorppi2018,
  author  = {Ockeloen-Korppi, C. F. and Damsk{\"a}gg, E. and Pirkkalainen, J.-M. and Asjad, M. and Clerk, A. A. and Massel, F. and Woolley, M. J. and Sillanp{\"a}{\"a}, M. A.},
  title   = {Stabilized entanglement of massive mechanical oscillators},
  journal = {Nature},
  volume  = {556},
  pages   = {478--482},
  year    = {2018},
  doi     = {10.1038/s41586-018-0038-x},
}

@article{Kotler2021,
  author  = {Kotler, Shlomi and Peterson, Gabriel A. and Shojaee, Ezad and Lecocq, Florent and Cicak, Katarina and Kwiatkowski, Alex and Geller, Shawn and Glancy, Scott and Knill, Emanuel and Simmonds, Raymond W. and Aumentado, Jos{\'e} and Teufel, John D.},
  title   = {Direct observation of deterministic macroscopic entanglement},
  journal = {Science},
  volume  = {372},
  pages   = {622--625},
  year    = {2021},
  doi     = {10.1126/science.abf2998},
}

@article{MercierdeLepinay2021,
  author  = {Mercier de L{\'e}pinay, Laure and Ockeloen-Korppi, Caspar F. and Woolley, Matthew J. and Sillanp{\"a}{\"a}, Mika A.},
  title   = {Quantum mechanics--free subsystem with mechanical oscillators},
  journal = {Science},
  volume  = {372},
  pages   = {625--629},
  year    = {2021},
  doi     = {10.1126/science.abf5389},
}

@article{Purdy2013,
  author  = {Purdy, T. P. and Yu, P.-L. and Kampel, N. S. and Peterson, R. W. and Clerk, A. A. and Regal, C. A.},
  title   = {Observation of radiation pressure shot noise on a macroscopic object},
  journal = {Science},
  volume  = {339},
  pages   = {801--804},
  year    = {2013},
  doi     = {10.1126/science.1231282},
}

@article{SafaviNaeini2013,
  author  = {Safavi-Naeini, Amir H. and Gr{\"o}blacher, Simon and Hill, Jeff T. and Chan, Jasper and Aspelmeyer, Markus and Painter, Oskar},
  title   = {Squeezed light from a silicon micromechanical resonator},
  journal = {Nature},
  volume  = {500},
  pages   = {185--189},
  year    = {2013},
  doi     = {10.1038/nature12307},
}

@article{Wollman2015,
  author  = {Wollman, E. E. and Lei, C. U. and Weinstein, A. J. and Suh, J. and Kronwald, A. and Marquardt, F. and Clerk, A. A. and Schwab, K. C.},
  title   = {Quantum squeezing of motion in a mechanical resonator},
  journal = {Science},
  volume  = {349},
  pages   = {952--955},
  year    = {2015},
  doi     = {10.1126/science.aac5138},
}

@article{Suh2014,
  author  = {Suh, J. and Weinstein, A. J. and Lei, C. U. and Wollman, E. E. and Steinke, S. K. and Meystre, P. and Clerk, A. A. and Schwab, K. C.},
  title   = {Mechanically detecting and avoiding the quantum fluctuations of a microwave field},
  journal = {Science},
  volume  = {344},
  pages   = {1262--1265},
  year    = {2014},
  doi     = {10.1126/science.1253258},
}

@article{Rossi2018,
  author  = {Rossi, Massimiliano and Mason, David and Chen, Jing and Tsaturyan, Yeghishe and Schliesser, Albert},
  title   = {Measurement-based quantum control of mechanical motion},
  journal = {Nature},
  volume  = {563},
  pages   = {53--58},
  year    = {2018},
  doi     = {10.1038/s41586-018-0643-8},
}

@article{Magrini2018,
  author  = {Magrini, Lorenzo and Rosenzweig, Philipp and Bach, Constanze and Deutschmann-Olek, Andreas and Hofer, Sebastian G. and Hong, Sungkun and Klocke, Niels and Aspelmeyer, Markus and Zeilinger, Anton and Groblacher, Simon},
  title   = {Real-time optimal quantum control of mechanical motion at room temperature},
  journal = {Nature},
  volume  = {595},
  pages   = {373--377},
  year    = {2021},
  doi     = {10.1038/s41586-021-03602-3},
}

@article{Rabl2011,
  author  = {Rabl, Peter},
  title   = {Photon blockade effect in optomechanical systems},
  journal = {Physical Review Letters},
  volume  = {107},
  pages   = {063601},
  year    = {2011},
  doi     = {10.1103/PhysRevLett.107.063601},
}

@article{Nunnenkamp2011,
  author  = {Nunnenkamp, A. and B{\o}rkje, K. and Girvin, S. M.},
  title   = {Single-photon optomechanics},
  journal = {Physical Review Letters},
  volume  = {107},
  pages   = {063602},
  year    = {2011},
  doi     = {10.1103/PhysRevLett.107.063602},
}

@article{Wootters1981,
  author  = {Wootters, William K.},
  title   = {Statistical distance and {H}ilbert-space geometric distance},
  journal = {Physical Review D},
  volume  = {23},
  pages   = {357--362},
  year    = {1981},
  doi     = {10.1103/PhysRevD.23.357},
}

@article{Giovannetti2004,
  author  = {Giovannetti, Vittorio and Lloyd, Seth and Maccone, Lorenzo},
  title   = {Quantum-enhanced measurements: beating the standard quantum limit},
  journal = {Science},
  volume  = {306},
  pages   = {1330--1336},
  year    = {2004},
  doi     = {10.1126/science.1104149},
}

@article{Giovannetti2006,
  author  = {Giovannetti, Vittorio and Lloyd, Seth and Maccone, Lorenzo},
  title   = {Quantum metrology},
  journal = {Physical Review Letters},
  volume  = {96},
  pages   = {010401},
  year    = {2006},
  doi     = {10.1103/PhysRevLett.96.010401},
}

@article{Boixo2007,
  author  = {Boixo, Sergio and Flammia, Steven T. and Caves, Carlton M. and Geremia, JM},
  title   = {Generalized limits for single-parameter quantum estimation},
  journal = {Physical Review Letters},
  volume  = {98},
  pages   = {090401},
  year    = {2007},
  doi     = {10.1103/PhysRevLett.98.090401},
}

@article{Wineland1994,
  author  = {Wineland, D. J. and Bollinger, J. J. and Itano, W. M. and Heinzen, D. J.},
  title   = {Squeezed atomic states and projection noise in spectroscopy},
  journal = {Physical Review A},
  volume  = {50},
  pages   = {67--88},
  year    = {1994},
  doi     = {10.1103/PhysRevA.50.67},
}

@article{Leibfried2005,
  author  = {Leibfried, D. and Barrett, M. D. and Schaetz, T. and Britton, J. and Chiaverini, J. and Itano, W. M. and Jost, J. D. and Langer, C. and Wineland, D. J.},
  title   = {Toward {H}eisenberg-limited spectroscopy with multiparticle entangled states},
  journal = {Science},
  volume  = {304},
  pages   = {1476--1478},
  year    = {2004},
  doi     = {10.1126/science.1097576},
}

@article{Huelga1997,
  author  = {Huelga, Susana F. and Macchiavello, Chiara and Pellizzari, Thomas and Ekert, Artur K. and Plenio, Martin B. and Cirac, J. Ignacio},
  title   = {Improvement of frequency standards with quantum entanglement},
  journal = {Physical Review Letters},
  volume  = {79},
  pages   = {3865--3868},
  year    = {1997},
  doi     = {10.1103/PhysRevLett.79.3865},
}

@article{Nagata2007,
  author  = {Nagata, Takafumi and Okamoto, Ryo and O'Brien, Jeremy L. and Sasaki, Keiji and Takeuchi, Shigeki},
  title   = {Beating the standard quantum limit with four-entangled photons},
  journal = {Science},
  volume  = {316},
  pages   = {726--729},
  year    = {2007},
  doi     = {10.1126/science.1138007},
}

@article{Afek2010,
  author  = {Afek, Ido and Ambar, Oron and Silberberg, Yaron},
  title   = {High {NOON} states by mixing quantum and classical light},
  journal = {Science},
  volume  = {328},
  pages   = {879--881},
  year    = {2010},
  doi     = {10.1126/science.1188172},
}

@article{Hyllus2012,
  author  = {Hyllus, Philipp and Laskowski, Wiesław and Krischek, Roland and Schwemmer, Christian and Wieczorek, Witlef and Weinfurter, Harald and Pezz{\'e}, Luca and Smerzi, Augusto},
  title   = {Fisher information and multiparticle entanglement},
  journal = {Physical Review A},
  volume  = {85},
  pages   = {022321},
  year    = {2012},
  doi     = {10.1103/PhysRevA.85.022321},
}

@article{Toth2012,
  author  = {T{\'o}th, G{\'e}za},
  title   = {Multipartite entanglement and high-precision metrology},
  journal = {Physical Review A},
  volume  = {85},
  pages   = {022322},
  year    = {2012},
  doi     = {10.1103/PhysRevA.85.022322},
}

@article{Demkowicz2009,
  author  = {Demkowicz-Dobrza{\'n}ski, Rafa{\l} and Dorner, Uwe and Smith, Brian J. and Lundeen, Jeff S. and Wasilewski, Wojciech and Banaszek, Konrad and Walmsley, Ian A.},
  title   = {Quantum phase estimation with lossy interferometers},
  journal = {Physical Review A},
  volume  = {80},
  pages   = {013825},
  year    = {2009},
  doi     = {10.1103/PhysRevA.80.013825},
}

@article{Schnabel2010,
  author  = {Schnabel, Roman and Mavalvala, Nergis and McClelland, David E. and Lam, Ping K.},
  title   = {Quantum metrology for gravitational wave astronomy},
  journal = {Nature Communications},
  volume  = {1},
  pages   = {121},
  year    = {2010},
  doi     = {10.1038/ncomms1122},
}

@article{Tse2019,
  author  = {Tse, M. and others},
  title   = {Quantum-enhanced advanced {LIGO} detectors in the era of gravitational wave astronomy},
  journal = {Physical Review Letters},
  volume  = {123},
  pages   = {231107},
  year    = {2019},
  doi     = {10.1103/PhysRevLett.123.231107},
}

@article{Acernese2019,
  author  = {Acernese, F. and others},
  title   = {Increasing the astrophysical reach of the advanced {VIRGO} detector},
  journal = {Physical Review Letters},
  volume  = {123},
  pages   = {231108},
  year    = {2019},
  doi     = {10.1103/PhysRevLett.123.231108},
}

@article{Bernad2018,
  author  = {Bern{\'a}d, J{\'o}zsef Zsolt and Sanavio, Claudio and Xuereb, Andr{\'e}},
  title   = {Optimal estimation of the optomechanical coupling strength},
  journal = {Physical Review A},
  volume  = {97},
  pages   = {063821},
  year    = {2018},
  doi     = {10.1103/PhysRevA.97.063821},
}

@article{Bernad2019,
  author  = {Bern{\'a}d, J{\'o}zsef Zsolt and Sanavio, Claudio and Xuereb, Andr{\'e}},
  title   = {Optimal estimation of matter-field coupling strength in the dipole approximation},
  journal = {Physical Review A},
  volume  = {99},
  pages   = {062106},
  year    = {2019},
  doi     = {10.1103/PhysRevA.99.062106},
}

@article{Schneiter2020,
  author  = {Schneiter, Fabienne and Qvarfort, Sofia and Serafini, Alessio and Xuereb, Andr{\'e} and Braun, Daniel and R{\"a}tzel, Dennis and Bruschi, David Edward},
  title   = {Optimal estimation with quantum optomechanical systems in the nonlinear regime},
  journal = {Physical Review A},
  volume  = {101},
  pages   = {033834},
  year    = {2020},
  doi     = {10.1103/PhysRevA.101.033834},
}

@article{Qvarfort2018,
  author  = {Qvarfort, Sofia and Serafini, Alessio and Barker, Peter F. and Bose, Sougato},
  title   = {Gravimetry through non-linear optomechanics},
  journal = {Nature Communications},
  volume  = {9},
  pages   = {3690},
  year    = {2018},
  doi     = {10.1038/s41467-018-06037-z},
}

@article{Sparaciari2015,
  author  = {Sparaciari, Caterina and Genoni, Matteo G. and Tufarelli, Tommaso and Adesso, Gerardo},
  title   = {Bounds to precision for quantum interferometry with {G}aussian states and operations},
  journal = {Journal of the Optical Society of America B},
  volume  = {32},
  pages   = {1354--1359},
  year    = {2015},
  doi     = {10.1364/JOSAB.32.001354},
}

@article{Wang2026,
  author  = {Wang, Yaqing and Sanavio, Claudio and Bern{\'a}d, J{\'o}zsef Zsolt},
  title   = {Optomechanical systems with linear and quadratic position couplings: dynamics and optimal estimation},
  journal = {Physical Review A},
  volume  = {113},
  pages   = {023504},
  year    = {2026},
  doi     = {10.1103/5dzx-dhql},
}

@article{Doeleman2023,
  author  = {Doeleman, Hendrik M. and Schatteburg, Thomas and Benevides, Rodrigo and Vollenweider, Samuel and Macri, Davide and Chu, Yiwen},
  title   = {Brillouin optomechanics in the quantum ground state},
  journal = {Physical Review Research},
  volume  = {5},
  pages   = {043140},
  year    = {2023},
  doi     = {10.1103/PhysRevResearch.5.043140},
}

@article{Fiaschi2021,
  author  = {Fiaschi, Niccol{\`o} and Hensen, Bas and Wallucks, Andreas and Benevides, Rodrigo and Li, Jie and Mayer Alegre, Thiago P. and Gr{\"o}blacher, Simon},
  title   = {Optomechanical quantum teleportation},
  journal = {Nature Photonics},
  volume  = {15},
  pages   = {817--821},
  year    = {2021},
  doi     = {10.1038/s41566-021-00866-z},
}

@article{Bild2023,
  author  = {Bild, Marius and Fadel, Matteo and Yang, Yu and von L{\"u}pke, Uwe and Martin, Phillip and Bruno, Alessandro and Chu, Yiwen},
  title   = {Schr{\"o}dinger cat states of a 16-microgram mechanical oscillator},
  journal = {Science},
  volume  = {380},
  pages   = {274--278},
  year    = {2023},
  doi     = {10.1126/science.adf7553},
}

@article{Ganapathy2023,
  author  = {Ganapathy, D. and Jia, W. and Nakano, M. and Xu, V. and Aritomi, N. and Cullen, T. and Kijbunchoo, N. and Dwyer, S. E. and Mullavey, A. and McCuller, L. and others},
  title   = {Broadband quantum enhancement of the {LIGO} detectors with frequency-dependent squeezing},
  journal = {Physical Review X},
  volume  = {13},
  pages   = {041021},
  year    = {2023},
  doi     = {10.1103/PhysRevX.13.041021},
}

@article{Backes2021,
  author  = {Backes, K. M. and Palken, D. A. and Al Kenany, S. and Brubaker, B. M. and Cahn, S. B. and Droster, A. and Hilton, Gene C. and Ghosh, Sumita and Jackson, H. and Lamoreaux, S. K. and others},
  title   = {A quantum enhanced search for dark matter axions},
  journal = {Nature},
  volume  = {590},
  pages   = {238--242},
  year    = {2021},
  doi     = {10.1038/s41586-021-03226-7},
}

@article{Fuderer2023,
  author  = {Fuderer, Liam A. and Hope, Joseph J. and Haine, Simon A.},
  title   = {Hybrid method of generating spin-squeezed states for quantum-enhanced atom interferometry},
  journal = {Physical Review A},
  volume  = {108},
  pages   = {043722},
  year    = {2023},
  doi     = {10.1103/PhysRevA.108.043722},
}

@article{LiaoLaw2011squeezing,
  author  = {Liao, Jie-Qiao and Law, C. K.},
  title   = {Parametric generation of quadrature squeezing of mirrors in cavity optomechanics},
  journal = {Physical Review A},
  volume  = {83},
  pages   = {033820},
  year    = {2011},
  doi     = {10.1103/PhysRevA.83.033820},
}

@article{LiaoLaw2011chirped,
  author  = {Liao, Jie-Qiao and Law, C. K.},
  title   = {Cooling of a mirror in cavity optomechanics with a chirped pulse},
  journal = {Physical Review A},
  volume  = {84},
  pages   = {053838},
  year    = {2011},
  doi     = {10.1103/PhysRevA.84.053838},
}

@article{LiaoCheungLaw2012,
  author  = {Liao, Jie-Qiao and Cheung, H. K. and Law, C. K.},
  title   = {Spectrum of single-photon emission and scattering in cavity optomechanics},
  journal = {Physical Review A},
  volume  = {85},
  pages   = {025803},
  year    = {2012},
  doi     = {10.1103/PhysRevA.85.025803},
}

@article{LiaoLaw2013,
  author  = {Liao, Jie-Qiao and Law, C. K.},
  title   = {Correlated two-photon scattering in cavity optomechanics},
  journal = {Physical Review A},
  volume  = {87},
  pages   = {043809},
  year    = {2013},
  doi     = {10.1103/PhysRevA.87.043809},
}

@article{LiaoLawKuangNori2015,
  author  = {Liao, Jie-Qiao and Law, C. K. and Kuang, Le-Man and Nori, Franco},
  title   = {Enhancement of mechanical effects of single photons in modulated two-mode optomechanics},
  journal = {Physical Review A},
  volume  = {92},
  pages   = {013822},
  year    = {2015},
  doi     = {10.1103/PhysRevA.92.013822},
}

@article{ZhouLiao2020,
  author  = {Zhou, Yue-Hui and Tan, Qing-Shou and Fang, Xi-Ming and Huang, Jin-Feng and Liao, Jie-Qiao},
  title   = {Spectrometric detection of weak forces in cavity optomechanics},
  journal = {Optics Express},
  volume  = {28},
  pages   = {28620--28634},
  year    = {2020},
  doi     = {10.1364/OE.398161},
}

\end{document}